\RequirePackage[displaymath]{lineno} 
\documentclass[aps,prd,twocolumn,showpacs,amsmath,amssymb]{revtex4-1}
\usepackage{epsfig}
\usepackage{graphicx}
\usepackage{dcolumn}
\usepackage{bm}
\usepackage{ltablex,booktabs}
\usepackage{overpic}
\usepackage{subfigure}
\usepackage{float}
\usepackage{color}
\usepackage{amsmath}
\usepackage{mathcomp}
\usepackage{mathrsfs}
\usepackage{multirow}
\usepackage{rotating}
\usepackage{amssymb}
\usepackage{gensymb}
\usepackage{amsmath}
\usepackage{tabularx}

\begin{document}
\normalsize
\parskip=5pt plus 1pt minus 1pt

\title{ \boldmath Amplitude Analysis and Branching Fraction Measurement of $D^{0} \rightarrow K^{-} \pi^{+} \pi^{0} \pi^{0}$ }
\vspace{-1cm}

\author{
\begin{small}
\begin{center}
M.~Ablikim$^{1}$, M.~N.~Achasov$^{10,d}$, S. ~Ahmed$^{15}$, M.~Albrecht$^{4}$, M.~Alekseev$^{55A,55C}$, A.~Amoroso$^{55A,55C}$, F.~F.~An$^{1}$, Q.~An$^{52,42}$, Y.~Bai$^{41}$, O.~Bakina$^{27}$
% ...
\\
\vspace{0.2cm}
(BESIII Collaboration)\\
\vspace{0.2cm} {\it
$^{1}$ Institute of High Energy Physics, Beijing 100049, People's Republic of China\\
% ...
$^{4}$ Bochum Ruhr-University, D-44780 Bochum, Germany\\
% ...
$^{10}$ G.I. Budker Institute of Nuclear Physics SB RAS (BINP), Novosibirsk 630090, Russia\\
% ...
$^{15}$ Helmholtz Institute Mainz, Johann-Joachim-Becher-Weg 45, D-55099 Mainz, Germany\\
% ...
$^{27}$ Joint Institute for Nuclear Research, 141980 Dubna, Moscow region, Russia\\
% ...
$^{41}$ Southeast University, Nanjing 211100, People's Republic of China\\
$^{42}$ State Key Laboratory of Particle Detection and Electronics, Beijing 100049, Hefei 230026, People's Republic of China\\
% ...
$^{52}$ University of Science and Technology of China, Hefei 230026, People's Republic of China\\
% ...
$^{55}$ (A)University of Turin, I-10125, Turin, Italy; (B)University of Eastern Piedmont, I-15121, Alessandria, Italy; (C)INFN, I-10125, Turin, Italy\\
% ...
$^{d}$ Also at the Novosibirsk State University, Novosibirsk, 630090, Russia\\
% ...
}\end{center}

\vspace{0.4cm}
\end{small}}

\affiliation{}
\vspace{-4cm}
\date{\today}
\begin{abstract}
Utilizing the dataset corresponding to an
integrated luminosity of $2.93$ fb$^{-1}$ at $\sqrt{s}=3.773$ GeV
collected by the BESIII detector, we report the first amplitude analysis
and branching fraction measurement of the $D^0\rightarrow K^-\pi^+\pi^0\pi^0$ decay.
We investigate the sub-structures and determine the relative fractions and 
the phases among the different intermediate processes.
Our results are used to provide an accurate detection efficiency and
allow measurement of
${\cal B}(D^0\rightarrow K^-\pi^+\pi^0\pi^0) \,=\,  
(8.86 \pm 0.13(\text{stat}) \pm 0.19(\text{syst}))\%$.     


\end{abstract}
\pacs{13.20.Fc, 12.38.Qk, 14.40.Lb}
\maketitle

\section{Introduction}
\label{sec:introduction}
Many measurements of $D$ meson decays have been performed since 
the $D$ mesons were discovered in 1976 
by Mark I~\cite{PhysRevLett.37.255, PhysRevLett.37.569}.  
Today, most of the low-multiplicity $D$ decay mode 
branching fractions (BFs) are well-measured. The largest decay modes are 
Cabibbo-favored (CF) hadronic (semileptonic) decay modes resulting 
from $c\rightarrow sW^+, W^+\rightarrow u\bar{d}\,(l^+\nu_l)$ transitions, 
but some of these decays are still unmeasured,
in which the $D^0\rightarrow K^-\pi^+\pi^0\pi^0$ decay
should be the largest unmeasured mode. 
Charge-conjugate states are implied throughout this paper.

The $D^0/D^+$ meson is the lightest meson containing a single charm quark. No 
strong decays are allowed, which makes the $D^0/D^+$ meson a perfect place to study the 
weak decay of the charm quark. 
The CF $\bar{K}\pi$, $\bar{K}2\pi$, and $\bar{K}3\pi$ modes are the 
most common hadronic decay modes of $D^0/D^+$ mesons. 
All $\bar{K}\pi$ and $\bar{K}2\pi$ branching fractions have been measured, 
but only four of the seven $\bar{K}3\pi$~\cite{KLModes} have been determined. 
The Mark III and The E691 collaboration performed amplitude analyses of all four 
$D\rightarrow \bar{K}\pi\pi\pi$ decay modes, 
$K^-\pi^+\pi^+\pi^-$, $K^0_S\pi^+\pi^+\pi^-$, 
$K^-\pi^+\pi^+\pi^0$, and $K^0_S\pi^+\pi^-\pi^0$~\cite{PhysRevD.45.2196, PhysRevD.46.1941}.
Recently, BESIII
has remeasured the structure of the $D^0\rightarrow K^-\pi^+\pi^+\pi^-$ decay with better precision~\cite{PhysRevD.95.072010}.
However, $\bar{K}3\pi$ modes with two or more $\pi^0$'s remain unmeasured. 

Furthermore, the $D^{0} \rightarrow K^{-}\pi^{+}\pi^{0}\pi^{0}$ decay
has a large BF and is often used as a $D^0$ meson 
``tag mode'' in experiment, such as in the CLEO and BESIII studies of 
$D^0$ semileptonic decays~\cite{PhysRevD.79.052010, PhysRevD.92.072012}. 
This mode contributes up to $10\%$ of the total reconstructed tags. 
Therefore, the accurate measurement of its sub-structures and branching 
fraction is essential to reduce systematic uncertainties of such analyses. 
While it is true that tag-mode BFs and sub-structure effects 
cancel to first order, higher-order systematic effects are increasingly 
important as statistics and precision increase. 


The BESIII detector collected a $2.93$ fb$^{-1}$ dataset in 2010 and 2011 at $\sqrt{s}=3.773$~GeV~\cite{1674-1137-37-12-123001, Ablikim2016629}, 
which corresponds to the mass of the $\psi(3770)$ resonance. The $\psi(3770)$ 
decays predominantly to $D^0\bar{D}^0$ or $D^+D^-$ without any additional hadrons.
The excellent tracking, precision calorimetry, and a large $D\bar{D}$ 
threshold data sample at BESIII provide an excellent opportunity for study 
of the unmeasured $D^0\rightarrow K^-\pi^+\pi^0\pi^0$ 
decay mode.  
The knowledge of intermediate structure will be crucial for determining the detection efficiency 
and useful for future usage as a tagging mode.
We report here the first partial wave analysis (PWA) and BF 
measurement of the $D^0\rightarrow K^-\pi^+\pi^0\pi^0$ decay.

\section{Detection and Data Sets}


The BESIII detector is described in detail in Ref.~\cite{ABLIKIM2010345}. The geometrical 
acceptance of the BESIII detector is 93\% of the full solid angle. Starting from the interaction 
point (IP), it consists of a main drift chamber (MDC), a time-of-flight (TOF) system, a CsI(Tl) 
electromagnetic calorimeter (EMC) and a muon system (MUC) with layers of resistive plate chambers 
(RPC) in the iron return yoke of a 1.0 T superconducting solenoid. The momentum resolution for 
charged tracks in the MDC is 0.5\% at a transverse momentum of 1~GeV/$c$.



Monte Carlo (MC) simulations of the BESIII detector are based on {\sc geant4}~\cite{sim}. 
The production of $\psi(3770)$ is simulated with the {\sc kkmc}~\cite{KKMC} package, taking into 
account the beam energy spread and the initial-state radiation (ISR). The {\sc photos}~\cite{FSR} 
package is used to simulate the final-state radiation of charged particles. The {\sc evtgen}~\cite{EvtGen} 
package is used to simulate the known decay modes with BFs taken from the Particle 
Data Group (PDG) ~\cite{PDG}, and the remaining unknown decays are generated with the {\sc LundCharm} 
model~\cite{LundCharm}. The MC sample referred to as “generic MC”, including the processes of $\psi(3770)$ 
decays to $D\bar{D}$, non-$D\bar{D}$, ISR production of low mass charmonium states and continuum 
($e^+e^- \rightarrow e^+e^-$, $\mu^+\mu^-$, $\gamma\gamma$ and $q\bar{q}$) processes, is used to study the background contribution. The effective 
luminosities of the generic MC sample correspond to at least 5 times the data luminosity.
The signal MC sample includes $D^0 \rightarrow K^- \pi^+ \pi^0 \pi^0$ versus $\bar{D}^0 \rightarrow K^+\pi^-$ 
events generated according to the results of the fit to data.

\section{Event Selection}
\label{chap:event_selection}
Photons are reconstructed as energy clusters in the EMC. The shower time is required 
be less than 700 ns from the event start time in order to suppress fake photons 
due to electronic noise or $e^+e^-$ beam background. Photon candidates within $\vert\cos\theta\vert < 0.80$ 
(barrel) are required to have larger than $25$ MeV energy deposition and those with 
$0.86<\vert\cos\theta\vert<0.92$ (endcap) must have larger than $50$ MeV energy deposition.
To suppress noise from hadronic shower splitoffs, the calorimeter positions 
of photon candidates must be at least $10\degree$ away from all charged tracks.

Charged track candidates from the MDC must satisfy $\vert\cos\theta\vert<0.93$, where $\theta$ 
is the polar angle with respect to the direction of the positron beam. The closest 
approach to the interaction point is required to be less than 10 cm in the beam 
direction and less than 1 cm in the plane perpendicular to the beam.

Charged tracks are identified as pions or kaons with particle identification (PID),
which is implemented by combining the information of $dE/dx$ in 
the MDC and the time-of-flight from the TOF system. For charged kaon 
candidates, the probability of the kaon hypothesis is required to 
be larger than that for a pion. For charged pion candidates, the 
probability for the pion  hypothesis is required to be larger than that for a kaon.

The $\pi^0$ candidates are reconstructed through $\pi^0\rightarrow \gamma\gamma$ decays,
with at least one barrel photon. The diphoton invariant mass is required to be in the range of 
$0.115<M_{\gamma\gamma}<0.150$~GeV/$c^2$. 

Two variables, beam constrained mass $M_{\text{BC}}$ and energy difference $\Delta E$, 
are used to identify the $D$ meson,
\begin{eqnarray}
\begin{aligned} 
M_{\text{BC}}&=\sqrt{E^2_{\text{beam}}-\vert\vec{p}_D\vert^2}\,,\\                                                                                                              
\Delta E&=E_D-E_{\text{beam}}\,,                                                                                                                                                
\end{aligned}
\end{eqnarray}
where $\vert\vec{p}_D\vert^2$ and $E_D$ are the total reconstructed momentum and 
energy of the $D$ candidate in the center-of-mass frame of the $\psi(3770)$, 
respectively, and $E_{\text{beam}}$ is the calibrated beam energy. The $D$ signals will 
be consistent with the nominal $D$ mass in $M_{\text{BC}}$ and with zero in 
$\Delta E$. 

After charged kaons and charged pions are identified, and neutral pions are reconstructed, hadronic $D$ 
decays can be reconstructed with a DTag technique. 
There are two types of samples used in the DTag technique: single tag (ST) and 
double tag (DT) samples. In the ST sample, only one $D$ or $\bar{D}$ meson is 
reconstructed through a chosen hadronic decay without any requirement on the remaining 
measured tracks and showers. For multiple ST candidates, only the candidate with the
smallest $\vert\Delta E\vert$ is kept. In the DT sample, both $D$ and $\bar{D}$ are 
reconstructed, where the meson reconstructed through the hadronic decay of interest 
is called the ``signal side'', and the other meson is called the ``tag side''.
For multiple DT candidates, only the candidate with the
smallest summation of $\vert\Delta E\vert$s in the signal side and the tag side is kept.

In this amplitude analysis, the DT candidates used are required to have the $D^0$ meson 
decaying to $K^- \pi^+ \pi^0 \pi^0$ as the signal, and the $\bar{D}^0$ meson decaying 
to $K^+\pi^-$ as the tag. For charged tracks of the signal side, a vertex fit is 
performed and the $\chi^2$ must be less than 100. To improve the resolution and ensure 
that all events fall within the phase-space boundary, we perform a three-constraint 
kinematic fit in which the invariant masses of the signal $D$ candidate and the two 
$\pi^0$'s are constrained to their PDG values~\cite{PDG}. The events with kinematic fit $\chi^2$ $>$ 
80 are discarded.

\begin{figure*}[hbtp]
\begin{center}
\centering
\begin{minipage}[b]{0.4\textwidth}
\epsfig{width=0.98\textwidth,file=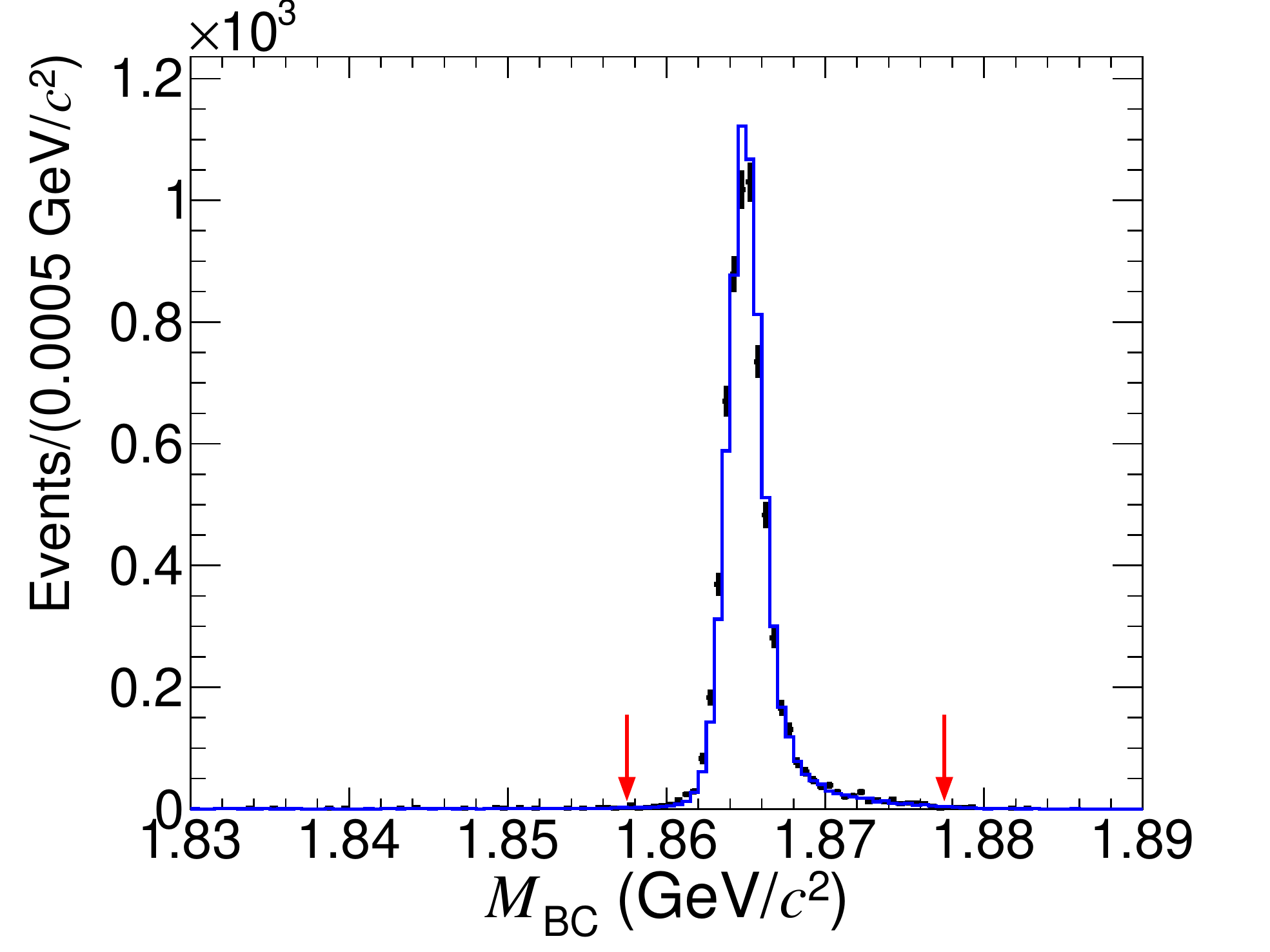}
\put(-50,100){(a)}
\end{minipage}
\begin{minipage}[b]{0.4\textwidth}
\epsfig{width=0.98\textwidth,file=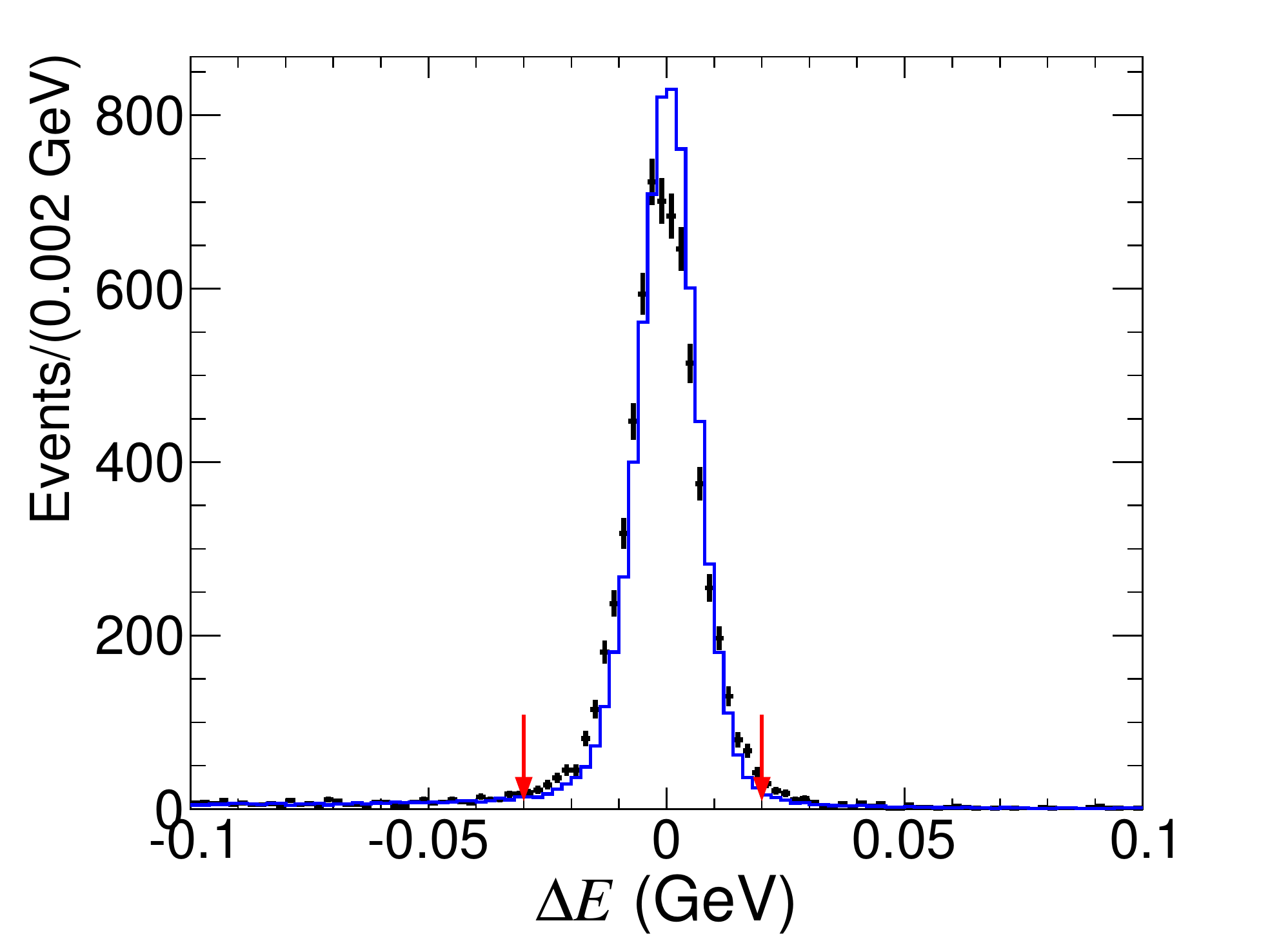}
\put(-50,100){(b)}
\end{minipage}
\begin{minipage}[b]{0.4\textwidth}
\epsfig{width=0.98\textwidth,file=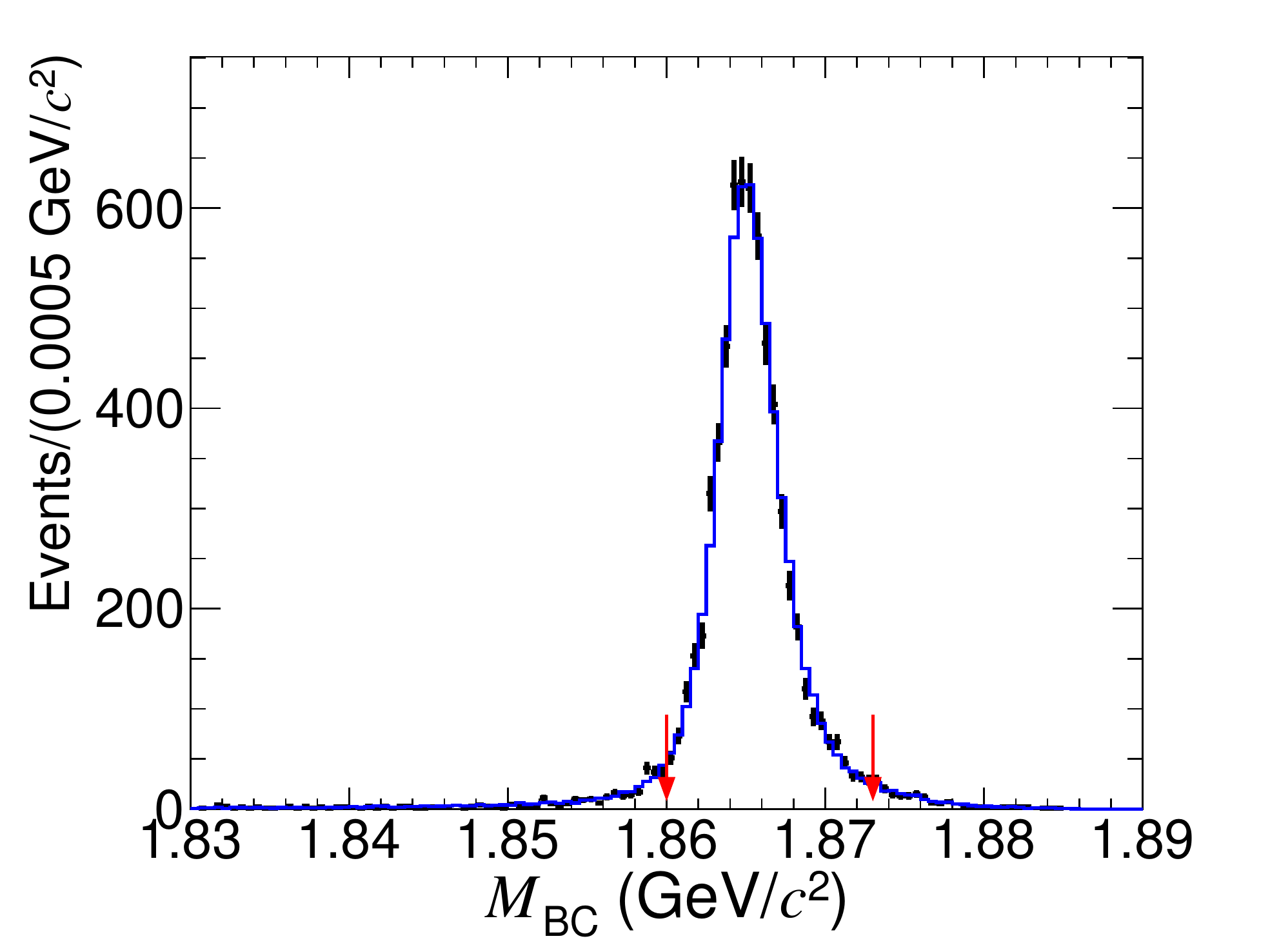}
\put(-50,100){(c)}
\end{minipage}
\begin{minipage}[b]{0.4\textwidth}
\epsfig{width=0.98\textwidth,file=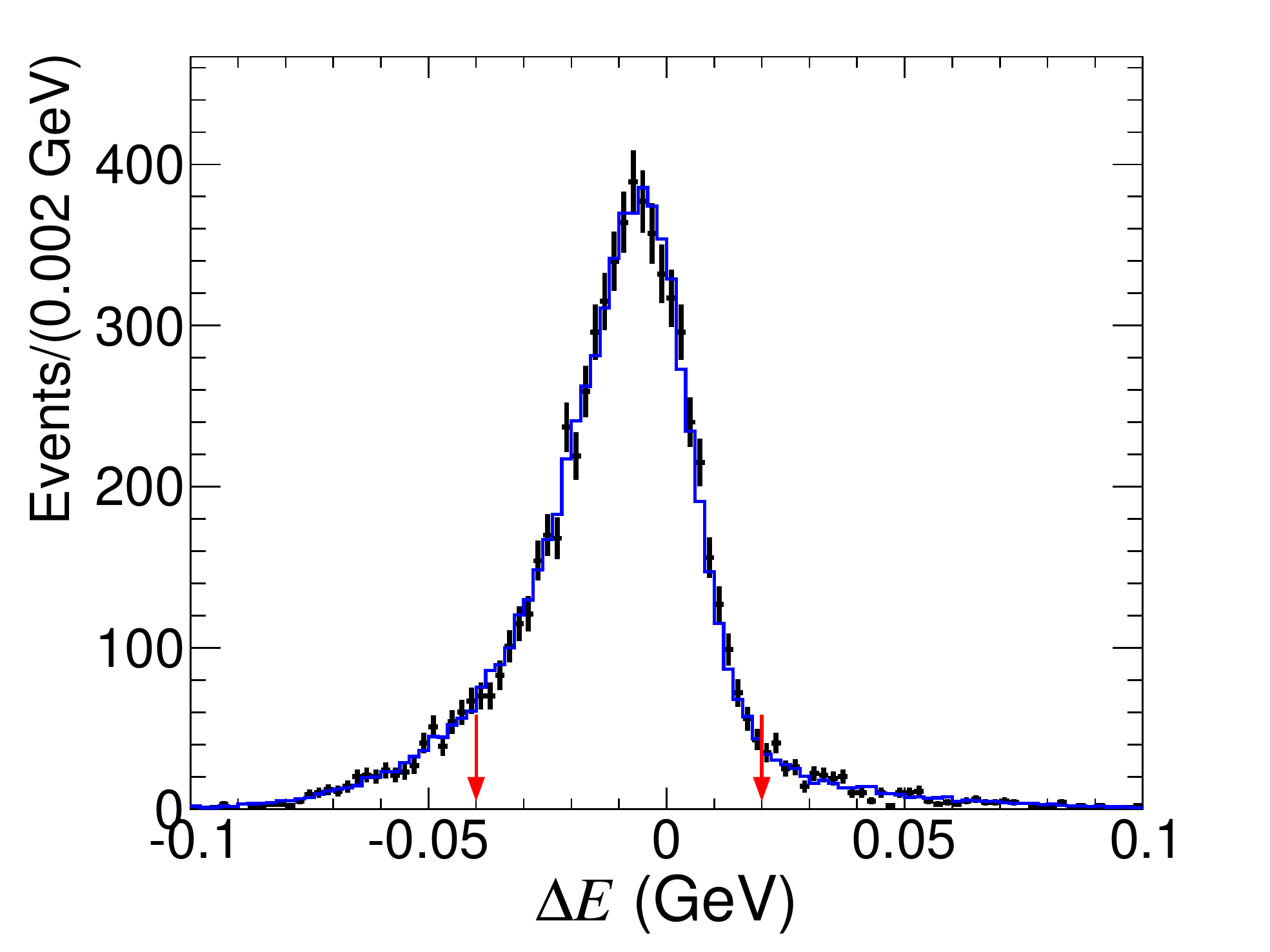}
\put(-50,100){(d)}
\end{minipage}
\caption{The (a) $M_{\text{BC}}$ and (b) $\Delta E$ distributions on the tag side. The (c) $M_{\text{BC}}$ and (d) $\Delta E$ distributions on the signal side. 
The (red) arrows indicate the requirements applied in the amplitude analysis. The (blue) solid lines indicate the MC sample, while the (black) dots with error bars indicate data.}
\label{fig:doubletag_mbc_deltaE_mode2}
\end{center}
\end{figure*}

The tag side is required to satisfy $1.8575<M_{\text{BC}}<1.8775$~GeV/$c^2$ and 
$-0.03<\Delta E<0.02$~GeV. 
The signal side is required to satisfy $1.8600<M_{\text{BC}}<1.8730$~GeV/$c^2$ and
$-0.04<\Delta E<0.02$~GeV.
A $K^0_S\rightarrow \pi^0\pi^0$ mass veto, 
$M_{\pi^0\pi^0}\notin (0.458, 0.520)$~GeV/$c^2$,
is also applied on the signal side to remove the 
dominant peaking background, $D^{0} \rightarrow K^-K^0_S\pi^+$.
The $M_{\text{BC}}$ and $\Delta E$ distributions of the data and generic MC samples 
are given in Fig.~\ref{fig:doubletag_mbc_deltaE_mode2},
where the generic MC sample is normalized to the size of data.
Note that we always apply the $\Delta E$ requirements before plotting $M_{\text{BC}}$, and vice-versa.

The generic MC sample is used to estimate the background of the DT candidates in the amplitude 
analysis. The dominant peaking background arises from $D^{0} \rightarrow K^{-}K^0_{S}\pi^{0}$, 
which is suppressed by the $K^0_S$ mass veto from $2.2\%$ to $0.07\%$.
The remaining non-peaking background is about 1.0$\%$. 
With all selection criteria applied, 5,950 candidate events are obtained with a purity of 98.9\%.


\section{Amplitude Analysis}
\label{Amplitude Analysis}
This analysis aims to determine the intermediate-state composition of the $D^0\to K^-\pi^+\pi^0\pi^0$ 
decay. This four-body decay spans a five-dimensional space. The daughter particle 
momenta are used as inputs to the probability density function (PDF) which 
describes the distribution of signal events.  This is then used in
an unbinned maximum likelihood fit to determinate the intermediate-state composition. 

\subsection{Likelihood Function Construction}
\label{sec:PDF_likelihood_fit}
The PDF is used to construct the likelihood of the amplitude mode and it is given by
\begin{eqnarray}
\begin{aligned} 
S(a,p) = \frac{\epsilon(p)\vert A(a,p)\vert^2R_4(p)}{\int\epsilon(p)\vert A(a,p)\vert^2R_4(p)dp}\,,                  
\end{aligned}
\end{eqnarray}
where $p$ is the set of the four daughter particles' four momenta and $a$ is the set of the complex 
coefficients for amplitude modes. The $\epsilon(p)$ is the efficiency parameterized 
in terms of the daughter particles' four momenta. The four-body phase-space, $R_4$, is  
defined as
\begin{eqnarray}
\begin{aligned} 
R_4(p)dp=(2\pi)^4\delta^4(p_D-\sum^4_\alpha p_\alpha)\prod^4_\alpha\frac{d^3p_\alpha}{(2\pi)^3 2E_\alpha}\,,           
\end{aligned}
\end{eqnarray}
where $\alpha$ indicates the four daughter particles.
This analysis uses an isobar model formulation, where
the signal decay amplitude, $A(a,p)$, is represented as a coherent sum of a number of two-body amplitude modes:
\begin{eqnarray}
\begin{aligned} 
\label{eq:total_amplitude}
A(a,p) = \sum_i a_iA_i(p)\,,            
\end{aligned}
\end{eqnarray}
where $a_i$ is written in the polar form as $\rho_i e^{i\phi_i}$ ($\rho_i$ is the magnitude and $\phi_i$ 
is the phase), and $A_i(p)$ is the amplitude for the $i^{\text{th}}$ amplitude mode modeled as
\begin{eqnarray}                                                                                                                                                                              
\begin{aligned}
A_i(p)=P^1_i(p)P^2_i(p)S_i(p)F^1_i(p)F^2_i(p)F^D_i(p)\,,                        
\end{aligned}                                                                                                                                                                                     
\end{eqnarray}
where the indexes 1 and 2 correspond to the two intermediate resonances.
Here, $F^{D}_i(p)$ is the Blatt-Weisskopf barrier factor for the $D$ meson, while
$P^{1,2}_i(p)$ and $F^{1,2}_i(p)$ are propagators and Blatt-Weisskopf barrier 
factors, respectively.
The spin factor 
$S_i(p)$ is constructed with the covariant tensor formalism~\cite{Zou2003}. 
Finally, the likelihood is defined as
\begin{eqnarray}                                                                                                                                                                              
\begin{aligned}
L=\prod_{k=1}^{N_s} S(a,p^k)\,, 
\end{aligned}                                                                                                                                                                                     
\end{eqnarray}
where $k$ sums over the selected events and $N_s$ is the number 
of candidate events. Consequently, the log likelihood is given by
\begin{eqnarray}                                                                                                                                                   
\begin{aligned} \label{eq:loglikelihood}                                                                                
\ln L&=\sum_{k=1}^{N_s} \ln S(a,p^k)\\                                                                         
     &=\sum_{k=1}^{N_s} \ln\left(\frac{\vert A(a,p^k)\vert^2}{\int\epsilon(p)\vert A(a,p)\vert^2R_4(p)dp} \right)\\                           
     &+\sum_{k=1}^{N_s}\ln R_4(p^k)+\sum_{k=1}^{N_s}\ln\epsilon(p^k)\,.    
\end{aligned}
\end{eqnarray}
Since the second term of Eq.~(\ref{eq:loglikelihood}) is independent of $a$
and the normalization integration in the denominator of the first term can be approximated 
by a phase-space MC integration, one can execute an amplitude analysis without knowing 
efficiency in advance. The phase-space MC integration is obtained by summing over a phase-space MC sample,
\begin{eqnarray}                                                                                                                                                               
\begin{aligned}
\label{eq:MCintegration}                                                                                    
&\int\epsilon(p)\vert A(a,p)\vert^2R_4(p)dp\\ 
&\approx \frac{1}{N_{\text{g,ph}}}\sum^{N_{\text{s,ph}}}_{l=1}\vert A(a,p^l)\vert^2\,,                                                                       
\end{aligned}
\end{eqnarray}
where $N_{\text{g,ph}}$ is the number of generated phase-space events and $N_{\text{s,ph}}$ is the 
number of selected phase-space events. 
This holds since the generated sample is uniform in phase space, 
while the nonuniform distribution after selection reflects the efficiency.


For signal MC samples, the amplitude squared for each event should be normalized by 
the PDF which generates the sample. The normalization integration using signal MC samples 
is given by
\begin{eqnarray}                                                                                                                                                                              
\begin{aligned}\label{eq:MCintegration_signal}                                                                                                                  
&\int\epsilon(p)\vert A(a,p)\vert^2R_4(p)dp\\  
&\approx \frac{1}{N_{\text{MC}}}\sum^{N_{\text{MC}}}_{l=1} \frac{\vert A(a,p^l)\vert^2}{\vert A(a^{\text{gen}},p^l)\vert^2}\,,                     
\end{aligned}
\end{eqnarray}                                                                                                                                                                              
where  $N_{\text{MC}}$ is the number of the signal MC sample and $a^{\text{gen}}$ 
is the set of the parameters used to generate the signal MC sample, which is obtained 
from the preliminary results using the phase-space MC integration. 
We allow for possible biases caused by tracking, PID,  and $\pi^0$ data versus MC sample efficiency differences by introducing 
the correction factors $\gamma_{\epsilon}$, 
\begin{eqnarray}                                                                                                                                                                              
\begin{aligned}
\label{eq:gamma_epsilon}                                                                                                                                                                          
\gamma_\epsilon(p)=\prod_j \frac{\epsilon_{j,\text{data}}(p)}{\epsilon_{j,\text{MC}}(p)}\,,               
\end{aligned}                                                                                                                                                                                     
\end{eqnarray}
where $\epsilon_{j,\text{data}}$ and $\epsilon_{j,\text{MC}}$ are the $\pi^0$ reconstruction, the PID, or the tracking efficiencies 
as a function of $p$ for the data and the MC sample, respectively. 
By weighting each signal MC event with $\gamma_{\epsilon}$, the MC integration is given by
\begin{eqnarray}                                                                                                                                                                              
\begin{aligned}\label{eq:MCintegration_signal_gamma}                                                                                                                                                             
&\int\epsilon(p)\vert A(a,p)\vert^2R_4(p)dp\\
&\approx \frac{1}{N_{\text{MC}}}\sum^{N_{\text{MC}}}_l \frac{\vert A(a,p^l)\vert^2 \gamma_{\epsilon}(p^l)}{\vert A(a^{\text{gen}},p^l)\vert^2}\,.                              
\end{aligned}                                                                                                                                                                                     
\end{eqnarray}

\subsubsection{Spin Factor}\label{ch:spin} 
For a decay process of the form
$a \rightarrow bc$, we use $p_a$, $p_b$, $p_c$ 
to denote the momenta of the particles $a$, $b$, $c$, respectively, and $r_a\,=\,p_b - p_c$. 
The spin projection operators~\cite{Zou2003} are defined as
\begin{eqnarray}                                                                                                                                                                              
\begin{aligned}
P^{(1)}_{\mu\mu^{\prime}}(a) &= -g_{\mu\mu^{\prime}}+\frac{p_{a,\mu}p_{a,\mu^{\prime}}}{p^2_a}\,,\\                                                                                                         
P^{(2)}_{\mu\nu\mu^{\prime}\nu^{\prime}}(a) &= \frac{1}{2}(P^{(1)}_{\mu\mu^{\prime}}(a)P^{(1)}_{\nu\nu^{\prime}}(a)+P^{(1)}_{\mu\nu^{\prime}}(a)P^{(1)}_{\nu\mu^{\prime}}(a))\\                         
&+\frac{1}{3}P^{(1)}_{\mu\nu}(a)P^{(1)}_{\mu^{\prime}\nu^{\prime}}(a)\,.                                                                                                                                         
\end{aligned}                                                                                                                                                                                     
\end{eqnarray}
The covariant tensors are given by
\begin{eqnarray}                                                                                                                                                                              
\begin{aligned}
\tilde{t}^{(1)}_{\mu}(a)&= -P^{(1)}_{\mu\mu^{\prime}}(a)r_a^{\mu^{\prime}}\,,\\                     
\tilde{t}^{(2)}_{\mu\nu}(a)&= P^{(2)}_{\mu\nu\mu^{\prime}\nu^{\prime}}(a)r_a^{\mu^{\prime}}r_a^{\nu^{\prime}}\,.                                                                                            
\end{aligned}                                                                                                                                                                                     
\end{eqnarray}
We list the 10 kinds of spin factors used in this analysis in Table~\ref{table:spin_factors}, 
where scalar, pseudo-scalar, vector, axial-vector, and tensor states are denoted 
by $S$, $P$, $V$, $A$, and $T$, respectively.
\begin{table*}[hbtp]
 \begin{center}
 \caption{Spin factor for each decay chain. All operators, i.e.~$\tilde{t}$, have the same definitions as Ref.~\cite{Zou2003}. 
   Scalar, pseudo-scalar, vector, axial-vector, and tensor states are denoted 
   by $S$, $P$, $V$, $A$, and $T$, respectively.}\label{table:spin_factors}
\begin{tabular}{ll}
 \hline
\hline
{\bf Decay chain}&{\bf $S(p)$} \\ 
 \hline
$D[S] \rightarrow V_1V_2$ & $\tilde{t}^{(1)\mu}(V_1) \; \tilde{t}^{(1)}_\mu(V_2)$ \\ 
$D[P] \rightarrow V_1V_2$ & $\epsilon_{\mu\nu\lambda\sigma}p^\mu(D) \; \tilde{T}^{(1)\nu}(D) \; \tilde{t}^{(1)\lambda}(V_1) \; \tilde{t}^{(1)\sigma}(V_2)$\\ 
$D[D] \rightarrow V_1V_2$ & $\tilde{T}^{(2)\mu\nu}(D) \; \tilde{t}^{(1)}_\mu(V_1) \; \tilde{t}^{(1)}_\nu(V_2)$\\ 
$D \rightarrow AP_1,A[S] \rightarrow VP_2$ & $\tilde{T}^{(1)\mu}(D) \; P^{(1)}_{\mu\nu}(A) \; \tilde{t}^{(1)\nu}{(V)}$ \\
$D \rightarrow AP_1,A[D] \rightarrow VP_2$ & $\tilde{T}^{(1)\mu}(D) \; \tilde{t}^{(2)}_{\mu\nu}(A)  \; \tilde{t}^{(1)\nu}(V)$ \\
$D \rightarrow AP_1,A \rightarrow SP_2$ & $\tilde{T}^{(1)\mu}(D) \; \tilde{t}^{(1)}_\mu(A)$ \\
$D \rightarrow VS$ & $\tilde{T}^{(1)\mu}(D) \; \tilde{t}^{(1)}_\mu(V)$ \\
$D \rightarrow V_1P_1,V_1\rightarrow V_2P_2$ & $\epsilon_{\mu\nu\lambda\sigma} \; p^\mu_{V_1} r^\nu_{V_1} \; p^\lambda_{P_1} \; r^\sigma_{V_2}$\\
$D \rightarrow PP_1,P\rightarrow VP_2$ & $p^\mu(P_2) \; \tilde{t}^{(1)}_\mu(V)$\\
$D \rightarrow TS$ & $\tilde{T}^{(2)\mu\nu}(D) \; \tilde{t}^{(2)}_{\mu\nu}(T)$\\
\hline
\hline
\end{tabular}
 \end{center}
\end{table*}

\subsubsection{Blatt-Weisskopf Barrier Factors}
The Blatt-Weisskopf barrier $F_i(p_j)$ is a barrier function for a two-body decay 
process, $a \rightarrow bc$. The Blatt-Weisskopf barrier depends on angular momenta 
and the magnitudes of the momenta of daughter particles in the rest system of the 
mother particle. The definition is given by
\begin{eqnarray}                                                                                                                                                                              
\begin{aligned}
F_L(q) = z^L X_L(q),                                                                                                                                                                               
\end{aligned}                                                                                                                                                                                     
\end{eqnarray}
where $L$ denotes the angular momenta, and $z\,=\,qR$ with $q$ the magnitudes 
of the momenta of daughter particles in the rest system of the mother particle and
$R$ the effective radius of the barrier. 
For a process $a\rightarrow bc$, we define $s_i = E^2_i-p_i^2$, $i=a,b,c$, such that
\begin{eqnarray}                                                                                                                                                                              
\begin{aligned}
q^2 = \frac{(s_a+s_b-s_c)^2}{4s_a}-s_b\,,
\end{aligned}                                                                                                                                                                                     
\end{eqnarray}
while the values of $R$ used in this analysis, $3.0$~GeV$^{-1}$ and $5.0$~GeV$^{-1}$ 
for intermediate resonances and the $D$ meson, respectively, are used in the BESIII MC generator
(based on {\sc evtgen}).
However, these values will also be varied as a source of systematic uncertainties. 
The $X_L(q)$ are given by
\begin{eqnarray}                                                                                                                                                                              
\begin{aligned}   
  X_{L=0}(q)&=1,\\   
  X_{L=1}(q)&=\sqrt{\frac{2}{z^2+1}},\\
  X_{L=2}(q)&=\sqrt{\frac{13}{9z^4+3z^2+1}}\,.
\end{aligned}                                                                                                                                                                                     
\end{eqnarray}

\subsubsection{Propagators}
We use the relativistic Breit-Wigner function as the propagator for the 
resonances $\bar{K}^{*0}$, $K^{*-}$, and $a_1(1260)^+$,
and fix their widths and masses to their PDG values~\cite{PDG}.
The relativistic Breit-Wigner function is given by
\begin{eqnarray}                                                                                                                                                                              
\begin{aligned}
P(m)=\frac{1}{(m^2_0-m^2)-im_0\Gamma(m)}\,,   
\end{aligned}                                                                                                                                                                                     
\end{eqnarray}
where $m=\sqrt{E^2-p^2}$ and $m_0$ is the rest mass of the resonance. $\Gamma(m)$ is given by
\begin{eqnarray}                                                                                                                                                                              
\begin{aligned}
\Gamma(m)=\Gamma_0\left(\frac{q}{q_0}\right)^{2L+1}\left(\frac{m_0}{m}\right)\left(\frac{X_L(q)}{X_L(q_0)}\right)^2\,,                                                                          
\end{aligned}                                                                                                                                                                                     
\end{eqnarray}
where $q_0$ indicates the value of $q$ when $s_a=m^2_0$.
Resonances $\bar{K}_1(1270)^{0}$ and $K_1(1270)^-$ are also parameterized
by the relativistic Breit-Wigner function but with constant width $\Gamma(m)=\Gamma_0$
since these two resonances are very close to the threshold of $\rho K$ and $\Gamma(m)$
vary very rapidly as $m$ changes.
We parameterize the $\rho$ with the Gounaris-Sakurai lineshape~\cite{PhysRevLett.21.244}, which is given by
\begin{eqnarray}                                                                                                                                                                              
\begin{aligned}
P_{\text{GS}}(m)=\frac{1+d\frac{\Gamma_0}{m_0}}{(m_0^2-m^2)+f(m)-im_0\Gamma(m)}\,.                                                                                                                                
\end{aligned}                                                                                                                                                                                     
\end{eqnarray}
The function $f(m)$ is given by
\begin{eqnarray}                                                                                                                                                                              
\begin{aligned}
&f(m)=\Gamma_0\frac{m_0^2}{q_0^3}\times\\
&\left[q^2(h(m)-h(m_0))+(m_0^2-m^2)q_0^2\left.\frac{dh}{d(m^2)}\right|_{m_0^2}\right]\,,     
\end{aligned}                                                                                                                                                                                     
\end{eqnarray}
where
\begin{eqnarray}                                                                                                                                                                              
\begin{aligned}
h(m)=\frac{2q}{\pi m}\ln\left(\frac{m+2q}{2m_{\pi}}\right)\,, 
\end{aligned}                                                                                                                                                                                     
\end{eqnarray}
and
\begin{align}  
&\left.\frac{dh}{d(m^2)}\right|_{m_0^2}\nonumber\\
&=h(m_0)\left[(8q_0^2)^{-1}-(2m_0^2)^{-1}\right]+(2\pi m_0^2)^{-1}\,.
\end{align}
The normalization condition at $P_{\text{GS}}(0)$ fixes the parameter $d=f(0)/(\Gamma_0 m_0)$. It is found to be
\begin{eqnarray}                                                                                                                                                                              
\begin{aligned}
d=\frac{3m^2_\pi}{\pi q_0^2}\ln\left(\frac{m_0+2q_0}{2m_\pi}\right)+\frac{m_0}{2\pi q_0}-\frac{m^2_\pi m_0}{\pi q^3_0}\,.  
\end{aligned}                                                                                                                                                                                     
\end{eqnarray}

\subsubsection{$K\pi$ $S$-Wave}\label{sec:kpi_swave}
The kinematic modifications associated with the $K\pi$ $S$-wave are modeled by a 
parameterization from scattering data~\cite{ASTON1988493,PhysRevD.78.034023}, which 
are described by a $K^{*0}$ Breit-Wigner along with an effective range non-resonant 
component with a phase shift,
\begin{eqnarray}\label{eq:kpi_swave_formfactor}                                                                                                                                    
\begin{aligned}
  A(m)=F\sin\delta_F e^{i\delta_F}+R\sin\delta_R e^{i\delta_R}e^{i2\delta_F}\,,                                                                                                                                   \end{aligned}
\end{eqnarray}
with
\begin{eqnarray}                                                                                                                                                                              
\begin{aligned}
  \delta_F&=\phi_F+\cot^{-1}\left[\frac{1}{aq}+\frac{rq}{2}\right]\nonumber\\                                                                                                                                     
  \delta_R&=\phi_R+\tan^{-1}\left[\frac{M\Gamma(m_{K\pi})}{M^2-m^2_{K\pi}}\right]\,,\nonumber                                                                                                                     \end{aligned}
\end{eqnarray}
where $a$ and $r$ are the scattering length and effective interaction length, respectively.
The parameters $F(\phi_F)$ and $R(\phi_R)$ are the magnitude (phase) for non-resonant state
and resonance terms, respectively. 
The parameters $M$, $F$, $\phi_F$,
$R$, $\phi_R$, $a$, $r$ are fixed to the results of the $D^0\rightarrow K^0_S\pi^+\pi^-$ analysis
by BABAR~\cite{PhysRevD.78.034023}, given in Table~\ref{tab:Kpi_swave_par}.
\begin{table}[ht]
 \caption{Parameters of $K\pi$ $S$-wave, by BABAR~\cite{PhysRevD.78.034023}, where the uncertainties 
include the statistical and systematic uncertainties.}\label{tab:Kpi_swave_par}
 \begin{center}
 \centering
\begin{tabular}{cc}
\hline
\hline
$M$(GeV/$c^2$) &$1.463\pm 0.002$\\
$\Gamma$(GeV) &$0.233\pm 0.005$\\ 
$F$ &$0.80\pm 0.09$\\ 
$\phi_F$&$2.33\pm0.13$\\
$R$&1(fixed)\\
$\phi_R$&$-5.31\pm 0.04$\\ 
$a$&$1.07\pm 0.11$\\ 
$r$&$-1.8\pm 0.3$\\ 
\hline
\hline
 \end{tabular}
 \end{center}
\end{table}
Note that we have also tested different parametrizations of the $\pi\pi$ $S$-wave, but no 
significant improvement is observed. We decide to use phase-space for the $\pi\pi$ $S$-wave. 

\subsection{Fit Fraction}\label{sec:fit_fraction}
The fit fraction (FF) is independent of the normalization and 
phase conventions in the amplitude formalism, and hence provides a more 
meaningful summary of amplitude strengths than the raw amplitudes, $\rho_i$ in Eq.~(\ref{eq:total_amplitude}), alone.
The definition of the FF for the $i^{\text{th}}$ amplitude is 
\begin{eqnarray}                                                                                                                                                                              
\begin{aligned}
\text{FF}_i &= \frac{\int\vert a_iA_i(p)\vert^2R_4(p)dp}{\int\vert\sum_k a_kA_k(p)\vert^2R_4(p)dp}\\                                                                            
     &\approx \frac{\sum_{l=1}^{N_{g,\text{ph}}}\vert a_iA_i(p^l)\vert^2}{\sum_{l=1}^{N_{g,\text{ph}}}\vert\sum_k a_kA_k(p^l)\vert^2}\,,
\end{aligned}
\end{eqnarray}
where the integration is approximated by a MC integration with a phase-space
MC sample. Since the FF does 
not involve efficiency, the MC sample used here is at the generator level 
instead of at the reconstruction level, 
as 
shown previously in Eq.~(\ref{eq:MCintegration}).

As for the statistical uncertainty of the FF, it is not practical to analytically 
propagate the uncertainties of the FFs from that of the amplitudes and phases. 
Instead, we randomly perturb the variables determined in our fit (by a Gaussian-distributed 
amount controlled by the fit uncertainty and the covariance matrix) and calculate 
the FFs to determine the statistical uncertainties. We fit the distribution 
of each FF with a Gaussian function and the width is reported 
as the uncertainty of the FF.

\subsection{Results of Amplitude Analysis}
We perform an unbinned likelihood fit using the likelihood described in Section 
\ref{sec:PDF_likelihood_fit}, where only the complex $a_i$ are floating.
Starting with amplitude modes with significant contribution, we add (remove) amplitude 
modes into (from) the fit one by one based on their statistical significances, 
which 
are obtained by the change of the log-likelihood value $\Delta\ln L$ 
with or without the amplitude mode under study.
There are 26 amplitudes each with a significance larger than $4\sigma$ chosen as the optimal set, listed 
in Table~\ref{tag:mode2_in} and the uncertainties are discussed in Section~\ref{sec:PWA_systematics}. There are more 
than 40 amplitudes tested but not used in the optimal set ($ < 4\sigma$ significance), listed in Appendix~A.

The amplitude $D\rightarrow K^-a_1(1260)^+$, $a_1(1260)^+\rightarrow\rho^+\pi^0[S]$ 
is expected to have the largest FF. Thus, we choose this amplitude 
as the reference (phase is fixed to 0) in the PWA. Other important amplitudes are
$D\rightarrow (K^-\pi^0)_S\rho^+$, $D\rightarrow K^-a_1(1260)^+$ with $a_1(1260)^+[S]\rightarrow\rho^+\pi^0$, and
$D\rightarrow K^-a_1(1260)^+$ with $a_1(1260)^+[S]\rightarrow\rho^+\pi^0$.
The notation $[S]$ denotes a relative $S$-wave between daughters in a decay, 
and similarly for $[P], [D]$.  
A MC sample is generated based on the PWA results, called the PWA signal MC sample.
The projections of the data sample and the PWA signal MC sample on the invariant masses squared and the cosines of helicity angles for 
the $K^-\pi^+$, $K^-\pi^0$, $\pi^+\pi^0$ and $\pi^0\pi^0$ systems are shown in Fig.~\ref{fig:PWA_fitting_projection}.  
The helicity angle $\theta_{ij}$ ($i$ or $j$ is $K^-$, $\pi^+$ 
and $\pi^0$) is defined as the angle between the momentum vector 
of the particle $i$ in the $ij$ rest frame and the direction of the $ij$ system 
in the $D$ rest frame.  
There are 
clear $K^*(892)^0$ and $K^*(892)^-$ resonances around 0.796~GeV$^2$$/c^4$ in the 
$M^2_{K^-\pi^+}$ and $M^2_{K^-\pi^0}$ projections, respectively, and a $\rho^+(770)$ 
resonance around 0.593~GeV$^2$$/c^4$ in the $M^2_{\pi^+\pi^0}$ projection. The gap 
in the $M^2_{\pi^0\pi^0}$ projection is due to the $K^0_S$ mass veto. 
A more detailed goodness-of-fit study is presented in the next section. 
The PWA signal MC sample
improves the accuracy of the DT efficiency (needed to determine 
the BF), which is discussed in more detail in Section~\ref{sec:efficiency}.

\subsection{Goodness-of-Fit}\label{sec:PWA_gof}
\begin{figure*}[!htp]
\centering
\begin{minipage}[b]{0.4\textwidth}
\epsfig{width=0.98\textwidth,file=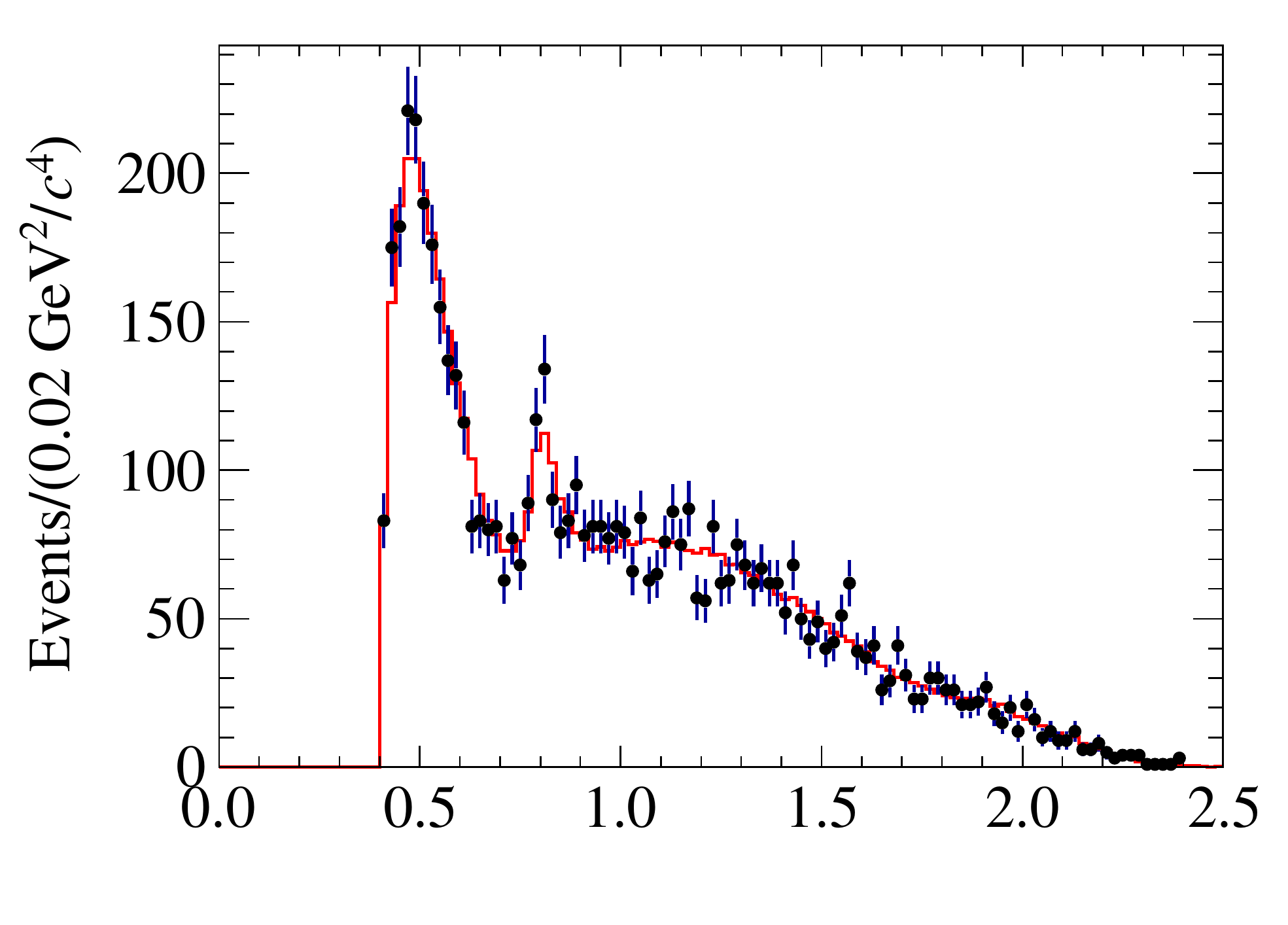}
\put(-110,3){\large $M^2_{K^-\pi^+}$ (GeV$^2$$/c^4$)}
\put(-30,120){(a)}
\end{minipage}
\begin{minipage}[b]{0.4\textwidth}
\epsfig{width=0.98\textwidth,file=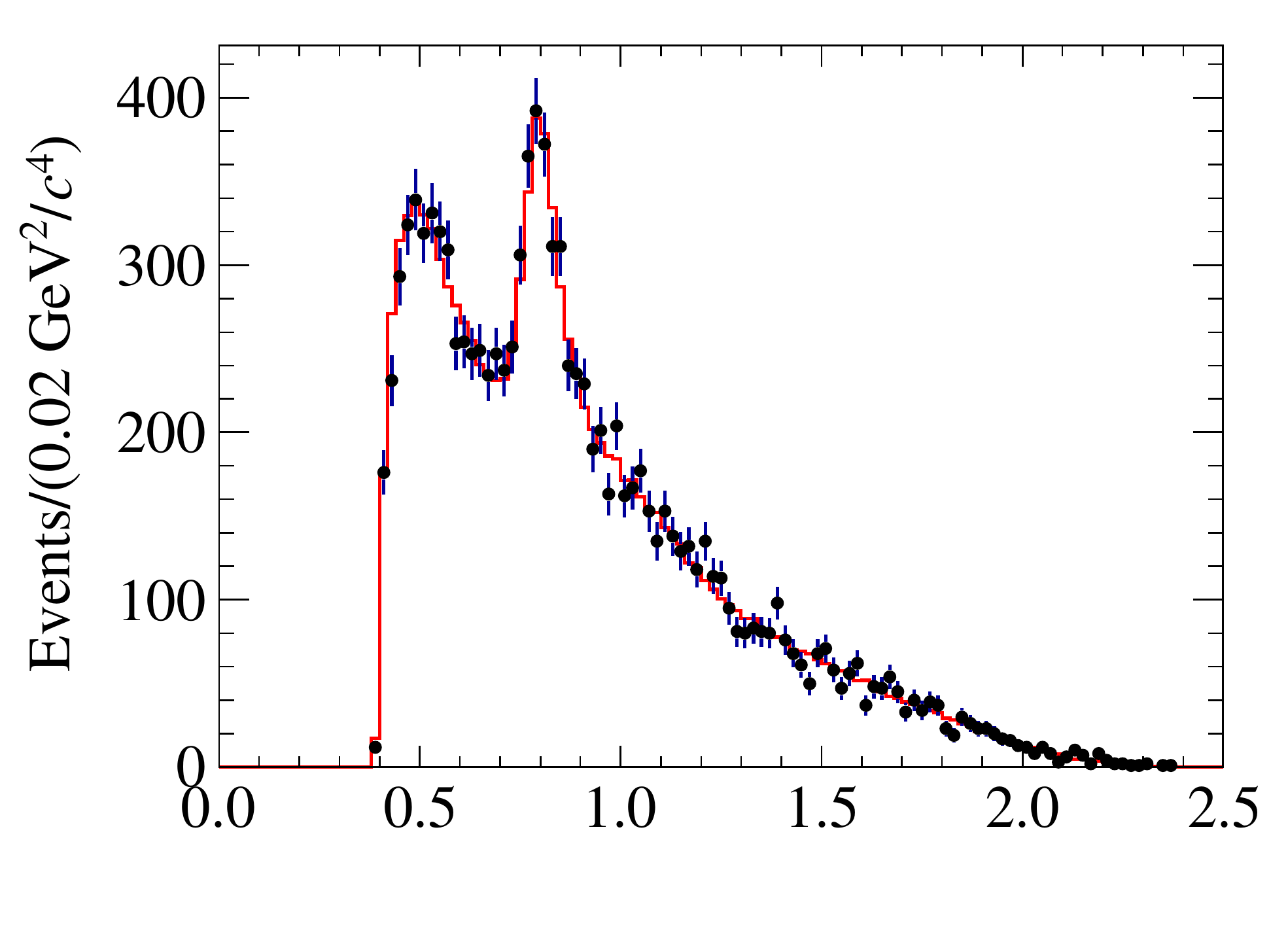}
\put(-110,3){\large $M^2_{K^-\pi^0}$ (GeV$^2$$/c^4$)}
\put(-30,120){(b)}
\end{minipage}
\begin{minipage}[b]{0.4\textwidth}
\epsfig{width=0.98\textwidth,file=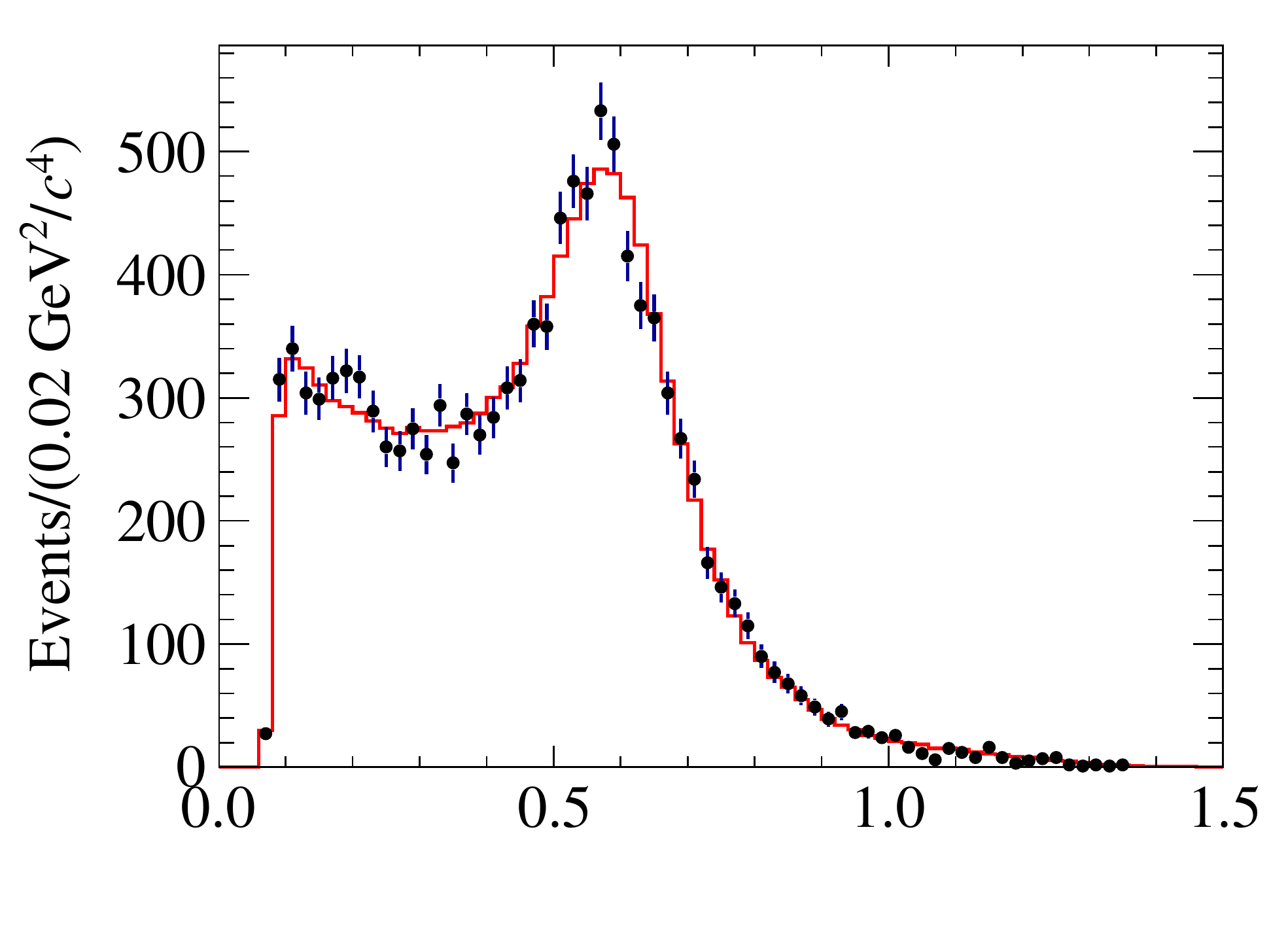}
\put(-110,3){\large $M^2_{\pi^+\pi^0}$ (GeV$^2$$/c^4$)}
\put(-30,120){(c)}
\end{minipage}
\begin{minipage}[b]{0.4\textwidth}
\epsfig{width=0.98\textwidth,file=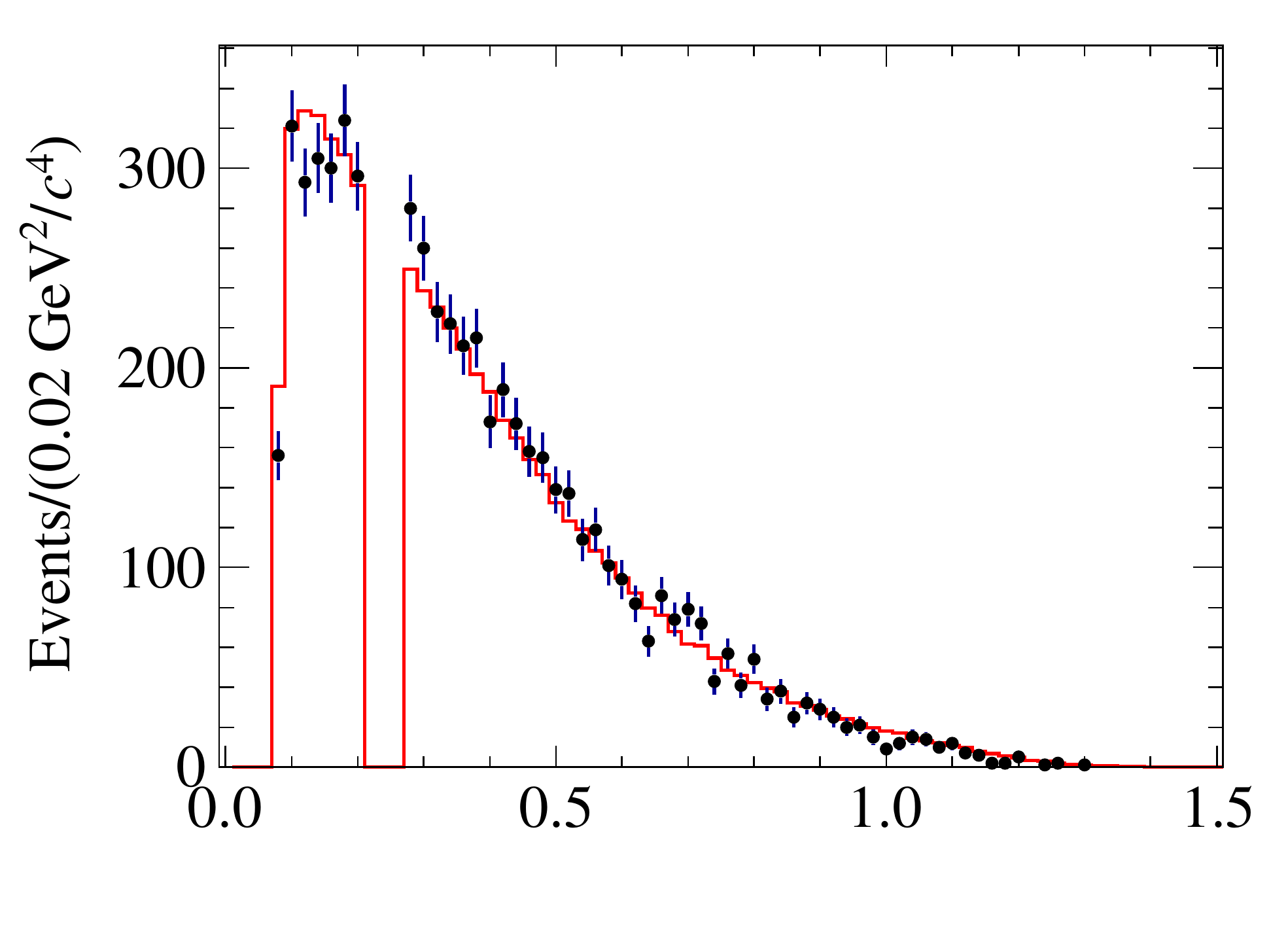}
\put(-110,3){\large $M^2_{\pi^0\pi^0}$ (GeV$^2$$/c^4$)}
\put(-30,120){(d)}
\end{minipage}
\begin{minipage}[b]{0.4\textwidth}
\epsfig{width=0.98\textwidth,file=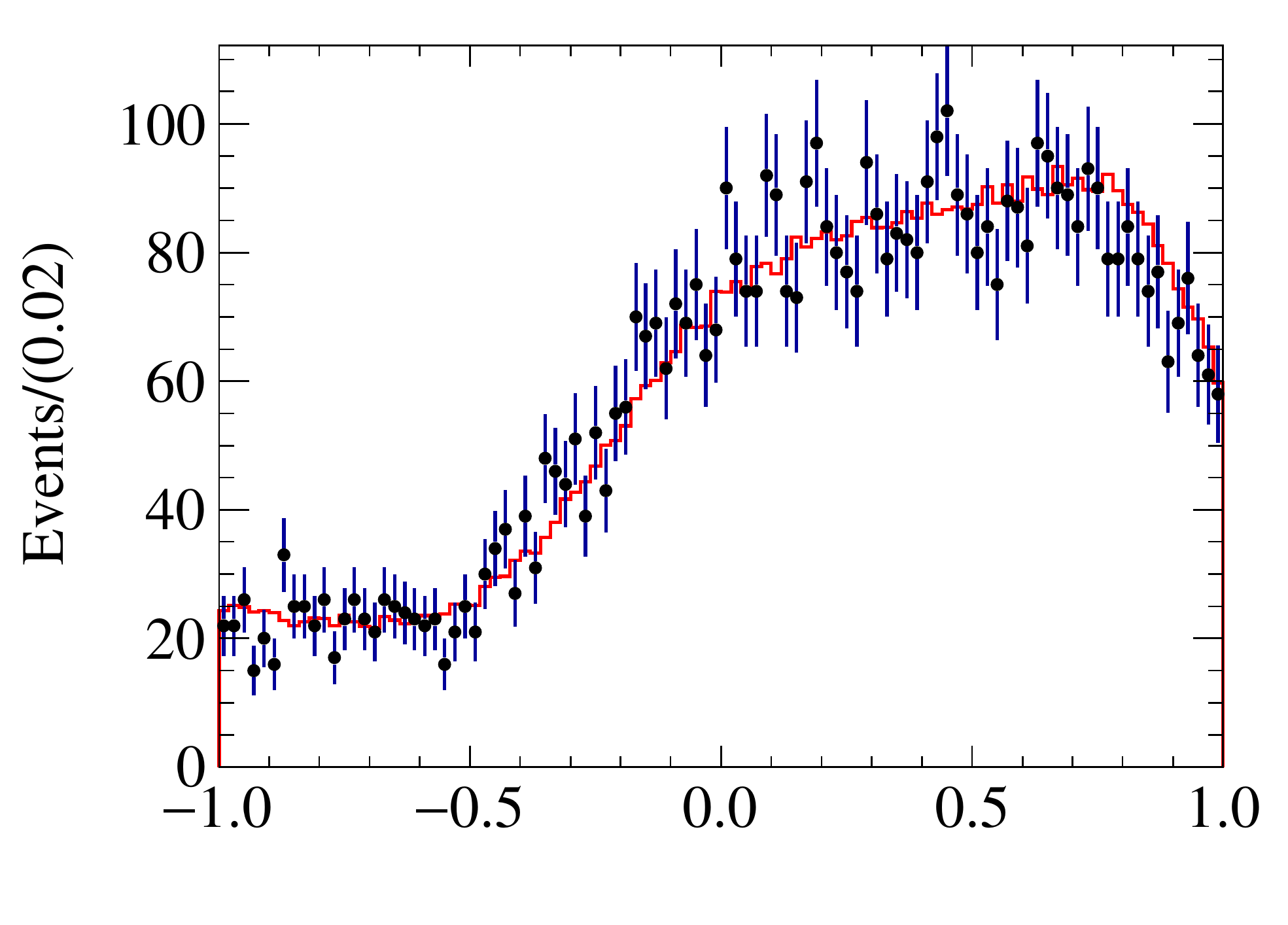}
\put(-100,3){\large $\cos\theta_{K^-\pi^+}$}
\put(-30,120){(e)}
\end{minipage}
\begin{minipage}[b]{0.4\textwidth}
\epsfig{width=0.98\textwidth,file=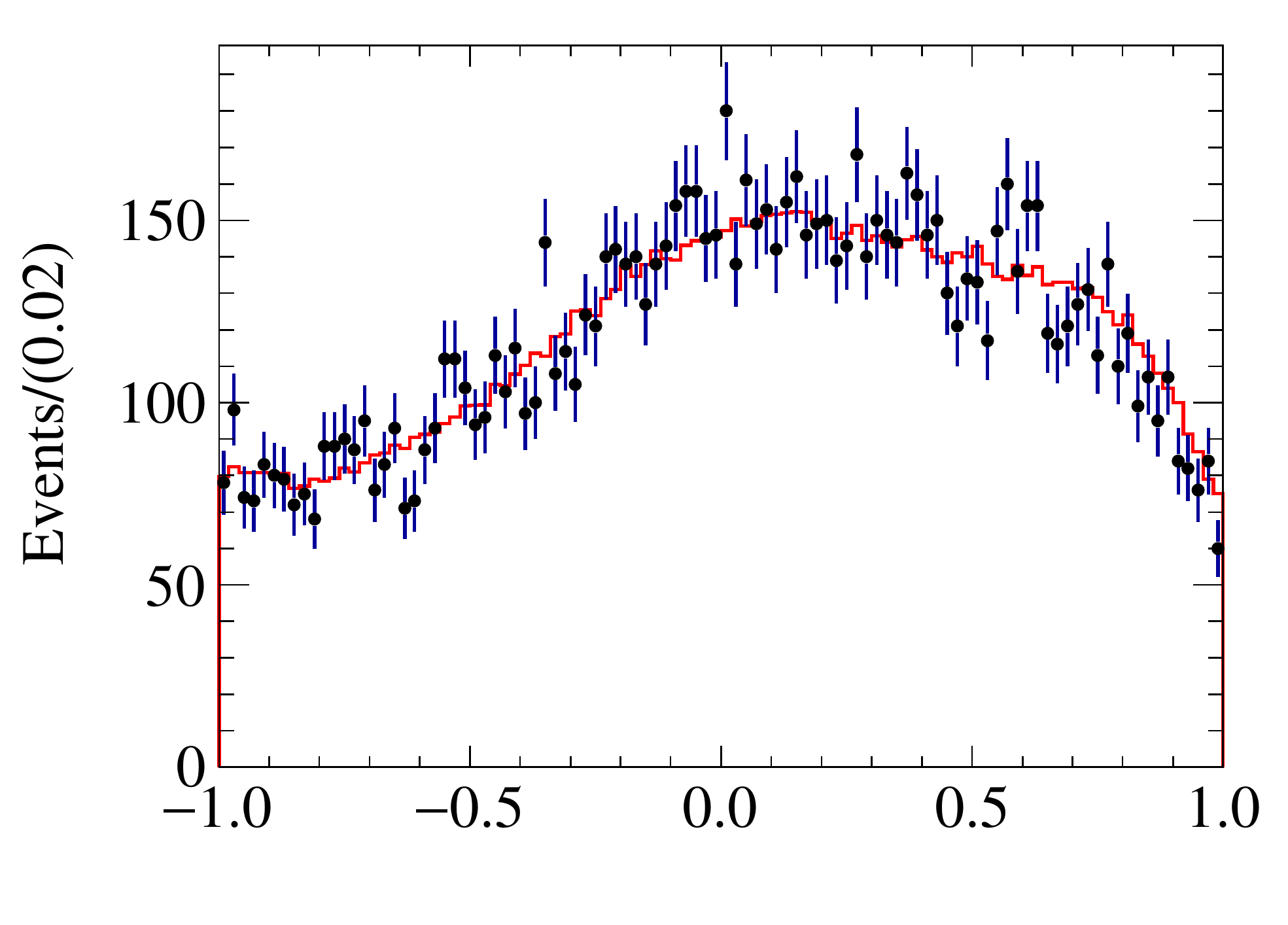}
\put(-100,3){\large $\cos\theta_{K^-\pi^0}$}
\put(-30,120){(f)}
\end{minipage}
\begin{minipage}[b]{0.4\textwidth}
\epsfig{width=0.98\textwidth,file=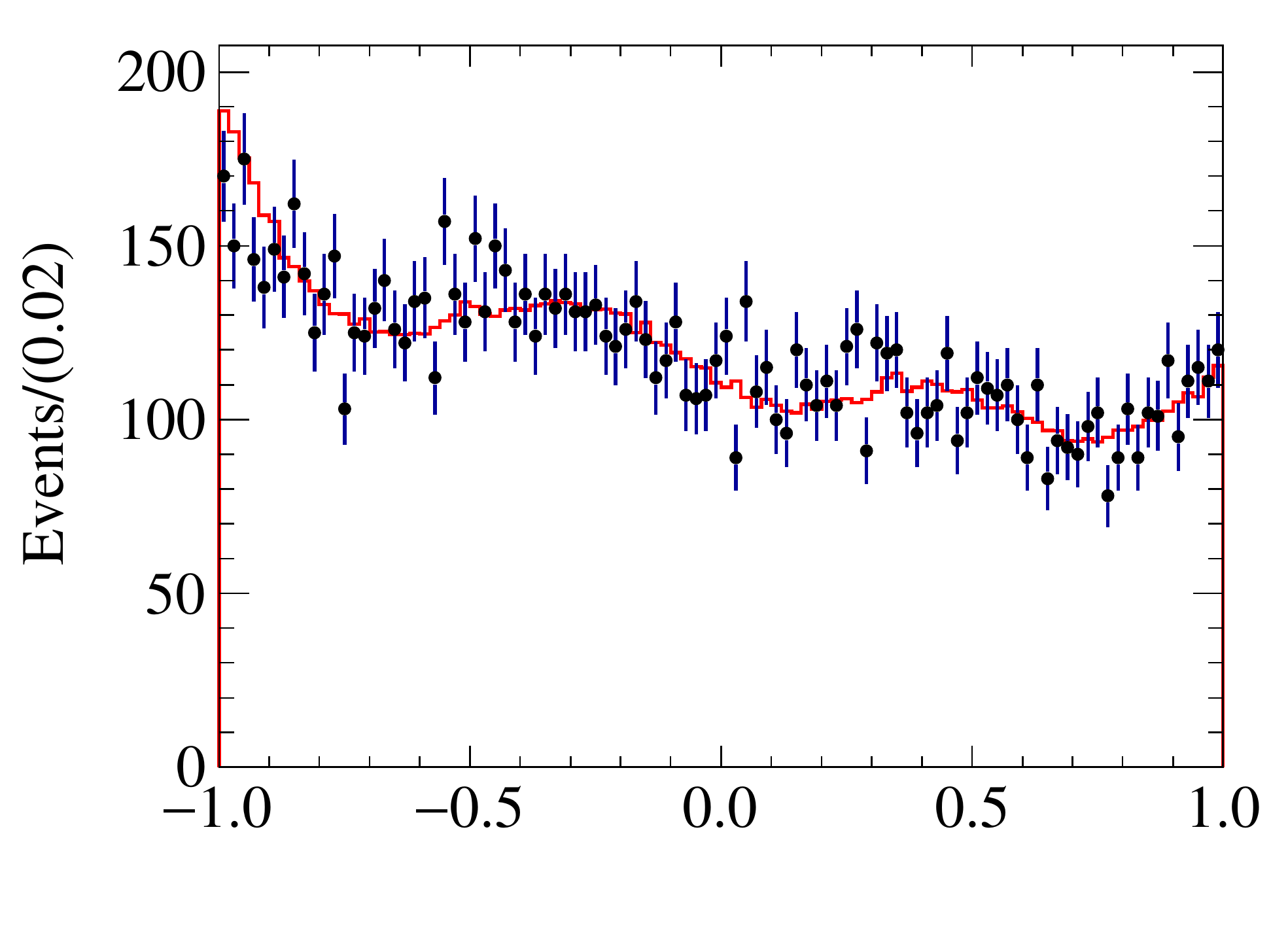}
\put(-100,3){\large $\cos\theta_{\pi^+\pi^0}$}
\put(-30,120){(g)}
\end{minipage}
\begin{minipage}[b]{0.4\textwidth}
\epsfig{width=0.98\textwidth,file=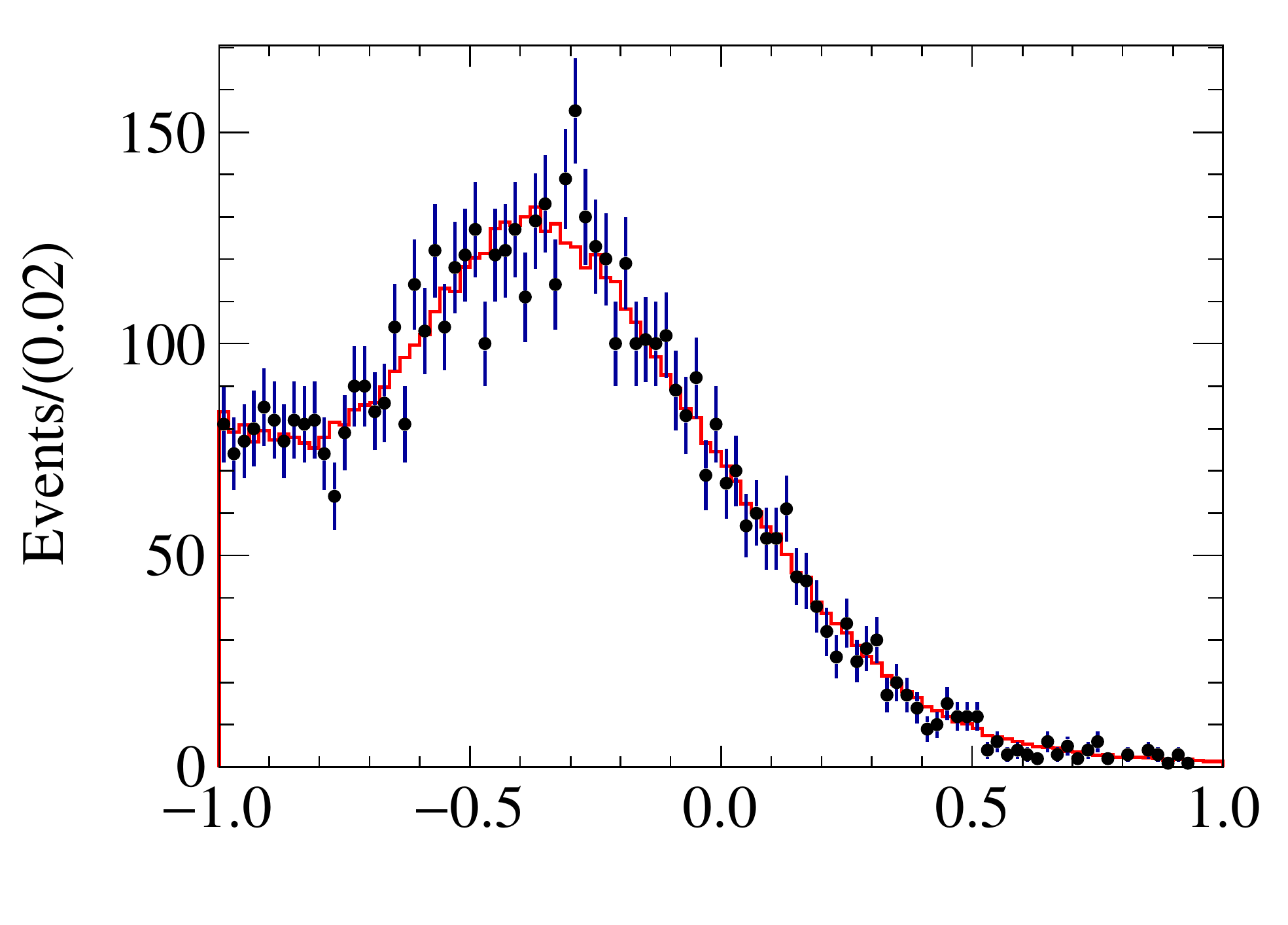}
\put(-100,3){\large $\cos\theta_{\pi^0\pi^0}$}
\put(-30,120){(h)}
\end{minipage}
\caption{Projections of the data sample and the PWA signal MC sample on the (a)-(d) invariant masses squared and the (e)-(h) cosines of helicity angles for the
$K^-\pi^+$, $K^-\pi^0$, $\pi^+\pi^0$ and $\pi^0\pi^0$ systems. The (red) solid lines indicate the fit results, while the (black) dots with error bars indicate data.} \label{fig:PWA_fitting_projection}
\end{figure*}
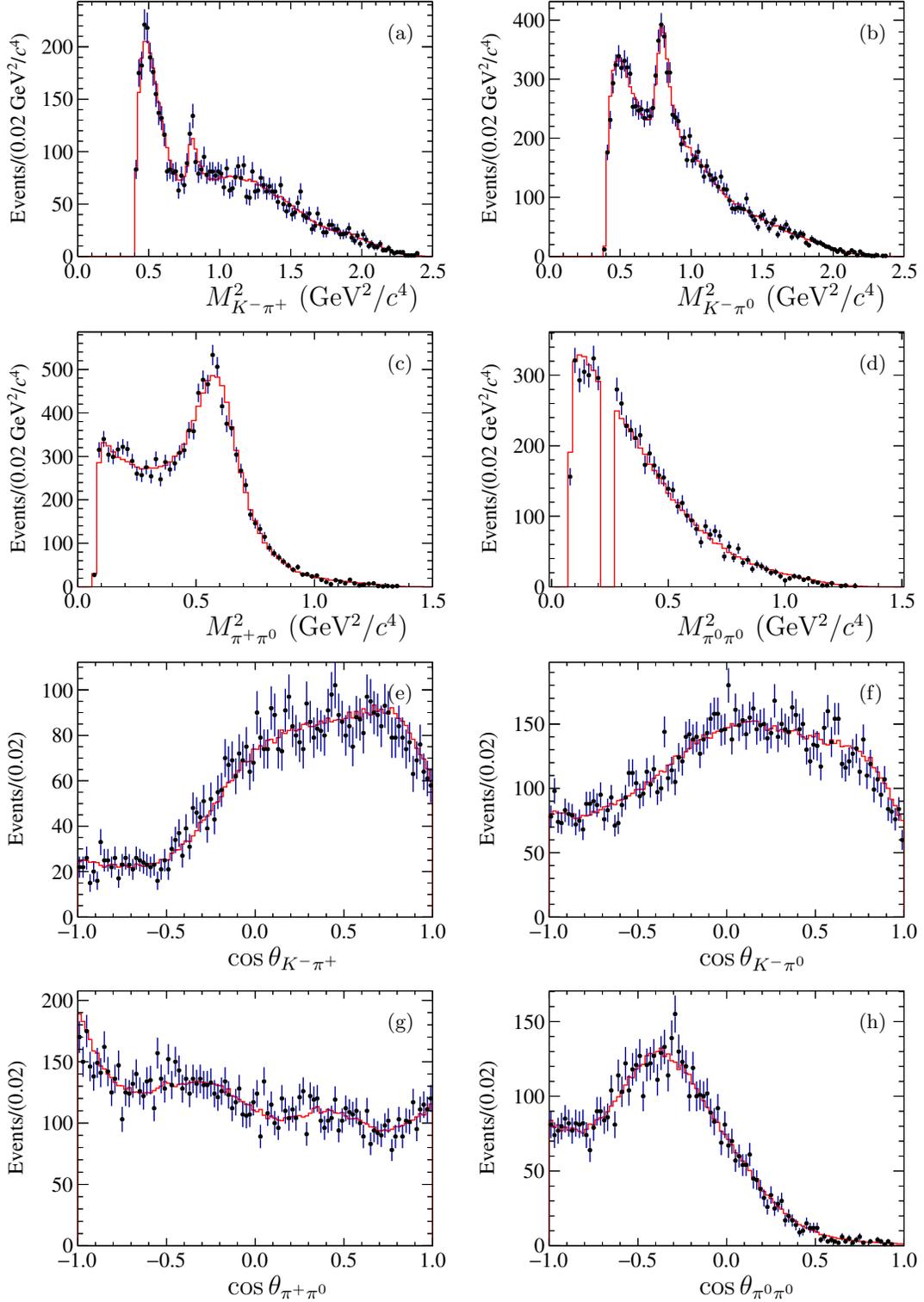

While the one-dimensional projections of the data sample and the PWA signal MC sample 
shown in Fig.~\ref{fig:PWA_fitting_projection}
look quite good, much information 
is lost in projecting down from the full five-dimensional phase space.  
It is thus desirable to have a more rigorous test of the fit quality.  
We have programmed a ``mixed-sample method'' for determining the goodness of our  
unbinned likelihood fit~\cite{1748-0221-5-09-P09004}. According to the method, 
we can calculate the ``T'' value of the mixing of two samples, the expectation 
mean, $\mu_\text{T}$, and the variance, $\sigma^2_\text{T}$. From these values, 
we can calculate a ``pull'',  $(\text{T}-\mu_\text{T})/\sigma_\text{T}$, 
which should distribute as a normal Gaussian function due to statistical fluctuations. 
The pull is expected to center at zero if the two samples come from the 
same parent PDF, and be biased toward larger values otherwise. In the case of 
our PWA fit, the pull is expected to be a little larger than zero because some 
amplitudes with small significance are dropped. In other words, adding more 
amplitudes into the model is expected to 
decrease the pull.

To check the goodness-of-fit of our PWA results, we calculate the pull of the T 
value of the mixing of the data sample and the PWA signal MC sample, 
and it is determined to be 0.97, which indicates good fit quality.  

\begin{table*}[ht]
 \caption[Amplitude modes included.]{FFs, phases, and significances of the optimal set of amplitude modes. 
The first and second uncertainties are statistical and systematic, respectively. 
The details of systematic uncertainties are discussed in Section~\ref{sec:PWA_systematics}.}\label{tag:mode2_in}
 \begin{center}
 \centering
\begin{tabular}{lllc}
 \hline
 \hline
{\bf Amplitude mode}&{\bf FF $(\%)$} &{\bf Phase $(\phi)$} &{\bf Significance $(\sigma)$}  \\\toprule
 \hline
$D\rightarrow SS$&& &\\
$D\rightarrow (K^-\pi^+)_{S\text{-wave}}(\pi^0\pi^0)_S$ & $ 6.92\pm 1.44\pm 2.86$&$          -0.75 \pm 0.15\pm 0.47$&$>10$\\
$D\rightarrow (K^-\pi^0)_{S\text{-wave}}(\pi^+\pi^0)_S$ & $ 4.18\pm 1.02\pm 1.77$&$          -2.90 \pm 0.19\pm 0.47$&$6.0$\\
 \hline
$D\rightarrow AP, A\rightarrow VP$&& &\\ 
$D\rightarrow K^-a_1(1260)^+,                                                                                                                                                                                  
\rho^+\pi^0[S]$                             & $28.36\pm 2.50\pm 3.53$&$\phantom{-}0$ (fixed)&$>10$\\ 
$D\rightarrow K^-a_1(1260)^+,                                                                                                                                                                                  
\rho^+\pi^0[D]$                             & $ 0.68\pm 0.29\pm 0.30$&$          -2.05 \pm 0.17\pm 0.25$&$6.1$\\ 
$D\rightarrow K_1(1270)^-\pi^+,                                                                                                                                                                                
K^{*-}\pi^0[S]$                             & $ 0.15\pm 0.09\pm 0.15$&$\phantom{-}1.84 \pm 0.34\pm 0.43$&$4.9$\\ 
$D\rightarrow K_1(1270)^0\pi^0,                                                                                                                                                                               
K^{*0}\pi^0[S]$                             & $ 0.39\pm 0.18\pm 0.30$&$          -1.55 \pm 0.20\pm 0.26$&$4.8$\\ 
$D\rightarrow K_1(1270)^0\pi^0,                                                                                                                                                                               
K^{*0}\pi^0[D]$                             & $ 0.11\pm 0.11\pm 0.11$&$          -1.35 \pm 0.43\pm 0.48$&$4.0$\\ 
$D\rightarrow K_1(1270)^0\pi^0,                                                                                                                                                                               
K^-\rho^+[S]$                               & $ 2.71\pm 0.38\pm 0.29$&$          -2.07 \pm 0.09\pm 0.20$&$>10$\\ 
$D\rightarrow (K^{*-}\pi^0)_A\pi^+,                                                                                                                                                                           
K^{*-}\pi^0[S]$                             & $ 1.85\pm 0.62\pm 1.11$&$\phantom{-}1.93 \pm 0.10\pm 0.15$&$7.8$\\ 
$D\rightarrow (K^{*0}\pi^0)_A\pi^0,                                                                                                                                                                           
K^{*0}\pi^0[S]$                             & $ 3.13\pm 0.45\pm 0.58$&$\phantom{-}0.44 \pm 0.12\pm 0.21$&$>10$\\ 
$D\rightarrow (K^{*0}\pi^0)_A\pi^0,                                                                                                                                                                           
K^{*0}\pi^0[D]$                             & $ 0.46\pm 0.17\pm 0.29$&$          -1.84 \pm 0.26\pm 0.42$&$5.9$\\ 
$D\rightarrow (\rho^+ K^-)_A\pi^0,                                                                                                                                                                            
K^-\rho^+[D]$                               & $ 0.75\pm 0.40\pm 0.60$&$\phantom{-}0.64 \pm 0.36\pm 0.53$&$5.1$\\ 
 \hline
$D\rightarrow AP, A\rightarrow SP$&& &\\ 
$D\rightarrow ((K^-\pi^+)_{S\text{-wave}}\pi^0)_A\pi^0$& $ 1.99\pm 1.08\pm 1.55$&$          -0.02 \pm 0.25\pm 0.53$&$7.0$\\ 
 \hline
$D\rightarrow VS$&& &\\ 
$D\rightarrow (K^-\pi^0)_{S\text{-wave}}\rho^+$        & $14.63\pm 1.70\pm 2.41$&$          -2.39 \pm 0.11\pm 0.35$&$>10$\\ 
$D\rightarrow K^{*-}(\pi^+\pi^0)_S$      & $ 0.80\pm 0.38\pm 0.26$&$\phantom{-}1.59 \pm 0.19\pm 0.24$&$4.1$\\ 
$D\rightarrow K^{*0}(\pi^0\pi^0)_S$      & $ 0.12\pm 0.12\pm 0.12$&$\phantom{-}1.45 \pm 0.48\pm 0.51$&$4.1$\\ 
 \hline
$D\rightarrow VP, V\rightarrow VP$&& &\\ 
$D\rightarrow (K^{*-}\pi^+)_V\pi^0$      & $ 2.25\pm 0.43\pm 0.45$&$\phantom{-}0.52 \pm 0.12\pm 0.17$&$>10$\\ 
 \hline
$D\rightarrow VV$&& &\\ 
$D\rightarrow K^{*-}\rho^+[S]$          & $ 5.15\pm 0.75\pm 1.28$&$\phantom{-}1.24 \pm 0.11\pm 0.23$&$>10$\\ 
$D\rightarrow K^{*-}\rho^+[P]$          & $ 3.25\pm 0.55\pm 0.41$&$          -2.89 \pm 0.10\pm 0.18$&$>10$\\ 
$D\rightarrow K^{*-}\rho^+[D]$          & $10.90\pm 1.53\pm 2.36$&$\phantom{-}2.41 \pm 0.08\pm 0.16$&$>10$\\ 
$D\rightarrow (K^-\pi^0)_V\rho^+[P]$   & $ 0.36\pm 0.19\pm 0.27$&$          -0.94 \pm 0.19\pm 0.28$&$5.7$\\ 
$D\rightarrow (K^-\pi^0)_V\rho^+[D]$   & $ 2.13\pm 0.56\pm 0.92$&$          -1.93 \pm 0.22\pm 0.25$&$>10$\\ 
$D\rightarrow K^{*-}(\pi^+\pi^0)_V[D]$ & $ 1.66\pm 0.52\pm 0.61$&$          -1.17 \pm 0.20\pm 0.39$&$7.6$\\ 
$D\rightarrow                                                                                                                                                                                              
(K^-\pi^0)_V(\pi^+\pi^0)_V[S]$                 & $ 5.17\pm 1.91\pm 1.82$&$          -1.74 \pm 0.20\pm 0.31$&$7.6$\\ 
 \hline
$D\rightarrow TS$&& &\\ 
$D\rightarrow (K^-\pi^+)_{S\text{-wave}}(\pi^0\pi^0)_T$& $ 0.30\pm 0.21\pm 0.30$&$          -2.93 \pm 0.31\pm 0.82$&$5.8$\\ 
$D\rightarrow (K^-\pi^0)_{S\text{-wave}}(\pi^+\pi^0)_T$& $ 0.14\pm 0.12\pm 0.10$&$\phantom{-}2.23 \pm 0.38\pm 0.65$&$4.0$\\ 
\hline
TOTAL&$98.54$&&\\ 
\hline
\hline
 \end{tabular}
 \end{center}
\end{table*}

\section{Branching Fraction}
We determinate the BF 
of $D^0 \rightarrow K^- \pi^+ \pi^0 \pi^0$ using the efficiency based on
the results of our amplitude analysis.
\subsection{Tagging Technique and Branching Fraction}
To extract the absolute BF of the $D^0 \rightarrow K^- \pi^+ \pi^0 \pi^0$ decay,
we obtain the ST sample by reconstructing the $\bar{D}^0$ meson through the $\bar{D}^0\rightarrow K^+\pi^-$ decay,
and the DT sample by fully reconstructing both $D^0$ and $\bar{D}^0$ through the $D^0\rightarrow K^-\pi^+\pi^0\pi^0$ decay
and the $\bar{D}^0\rightarrow K^+\pi^-$ decay as the signal side and the tag side, respectively. 
The ST yield is given by
\begin{eqnarray}
\begin{aligned}
\label{eq:ST_yield}                                                                                                                                                                               
N^{\text{ST}}_{\text{tag}} = 2N_{D^0\bar{D}^0}{\cal B}_{\text{tag}}\varepsilon_{\text{tag}}\,,                                                                                                       
\end{aligned}
\end{eqnarray}
and the DT yield is given by
\begin{eqnarray}
\begin{aligned}
\label{eq:DT_yield}                                                                                                                                                                            
N^{\text{DT}}_{\text{tag,sig}} = 2N_{D^0\bar{D}^0}{\cal B}_{\text{tag}}{\cal B}_{\text{sig}}\varepsilon_{\text{tag,sig}}\,,                                                           
\end{aligned}
\end{eqnarray}
where $N_{D^0\bar{D}^0}$ is the total number of produced $D^0\bar{D}^0$ pairs,
${\cal B}_{\text{tag(sig)}}$ is the BF of the tag (signal) side,
and $\varepsilon$ are the corresponding efficiencies.

The BF of the signal side is determined by isolating 
${\cal B}_{\text{sig}}$ such that
\begin{eqnarray}
\begin{aligned}
\label{eq:DTag_BF}                                                                                                                                                                          
{\cal B}_{\text{sig}} = \frac{N^{\text{DT}}_{\text{tag,sig}}}{N^{\text{ST}}_{\text{tag}}}\frac{\varepsilon_{\text{tag}}}{\varepsilon_{\text{tag,sig}}}\,.                                                 
\end{aligned}
\end{eqnarray}
\subsection{Fitting Model}
The ST yield, $N^{\text{ST}}_{\text{tag}}$, is obtained by a maximum-likelihood fit 
to the $M_{\text{BC}}$ ($K^+\pi^-$) distribution. A Crystal Ball (CB) function~\cite{CBcitation}, 
along with a Gaussian, is used to model the signal while an ARGUS function~\cite{ARGUScitation} is used 
to model the background. The signal shape is
\begin{eqnarray}
\begin{aligned}
f\times\text{CB}(x;\mu,\sigma,\alpha,n)+(1-f)\text{Gaussian}(\mu_G,\sigma_G)\,,  
\end{aligned}
\end{eqnarray}
where $f$ is a fraction ranging from 0 to 1, $\mu_G$ and $\sigma_G$ are the mean and 
the width of the Gaussian function, respectively. The CB function has a 
Gaussian core transitioning to a power-law tail at a certain point, and is given by
\begin{align}                                                                                                                                                                                                  
&\text{CB}(x;\mu,\sigma,\alpha,n)
=N\times\nonumber\\ &
\left\{\begin{array}{@{}ll@{}}                                                                                                                                                                                   
    \text{exp}\left(-\frac{(x-\mu)^2}{2\sigma^2}\right)\,, & \text{if}\ \frac{x-\mu}{\sigma}>\alpha \\                                                                                                            
    \left(\frac{n}{\vert\alpha\vert}\right)^n e^{\frac{-\vert\alpha\vert^2}{2}}                                                                                                                 
\left(\frac{n-\vert\alpha\vert^2}{\vert\alpha\vert}-\frac{x-\mu}{\sigma}\right)^{-n} & \text{otherwise}                                                                                              
  \end{array}\right.   
\end{align}
where $N$ is the normalization and $\alpha$ controls the start of the tail.
The beam energy (end point of the ARGUS function) is fixed to be 1.8865~GeV,
while all other parameters are floating.

The DT yield, $N^{\text{DT}}_{\text{tag,sig}}$, is obtained by a maximum-likelihood 
fit to the two dimensional (2-D) $M_{\text{BC}}$ ($K^-\pi^+\pi^0\pi^0$) versus $M_{\text{BC}}$ ($K^+\pi^-$) 
distribution for the signal and the tag side with a 2-D fitting technique 
introduced by CLEO~\cite{PhysRevD.76.112001}. This technique analytically models 
the signal peak, and considers ISR and mispartition 
(i.e., where one or more daughter particles are associated with 
   the incorrect $D^0$ or $\bar{D}^0$ parent) effects,
which are non-factorizable in the 2-D plane. In this fitting, the mass of $\psi(3770)$ 
is fixed to be 3.773~GeV and the beam energy is fixed to be 1.8865~GeV.

\subsection{Efficiency and Data Yields}\label{sec:efficiency}
\begin{figure*}[!htp]
\begin{center}
\centering
\begin{minipage}[b]{.62\columnwidth}
\epsfig{width=0.98\textwidth,file=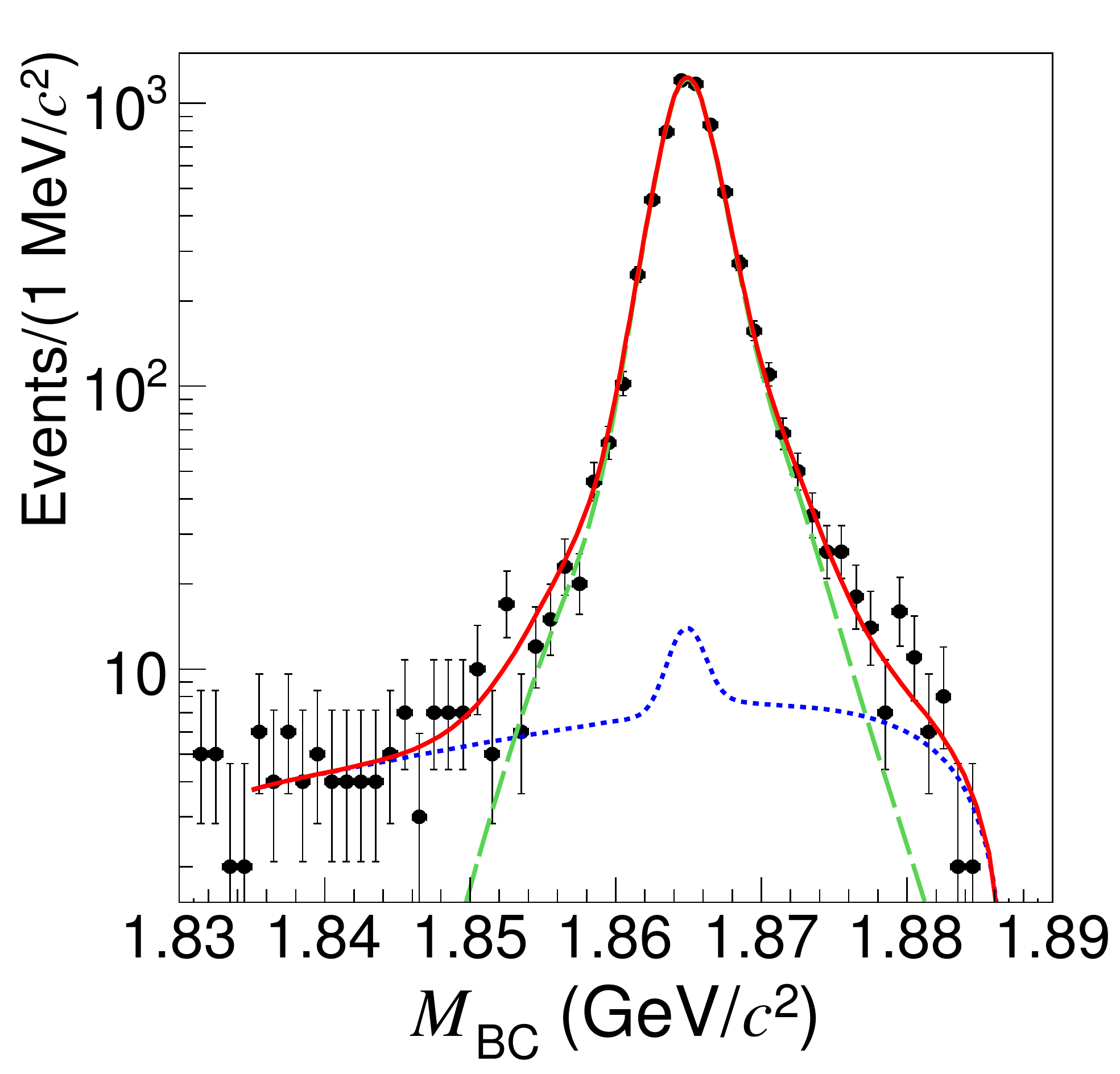}
\put(-25,120){(a)}
\end{minipage}
\begin{minipage}[b]{.62\columnwidth}
\epsfig{width=0.98\textwidth,file=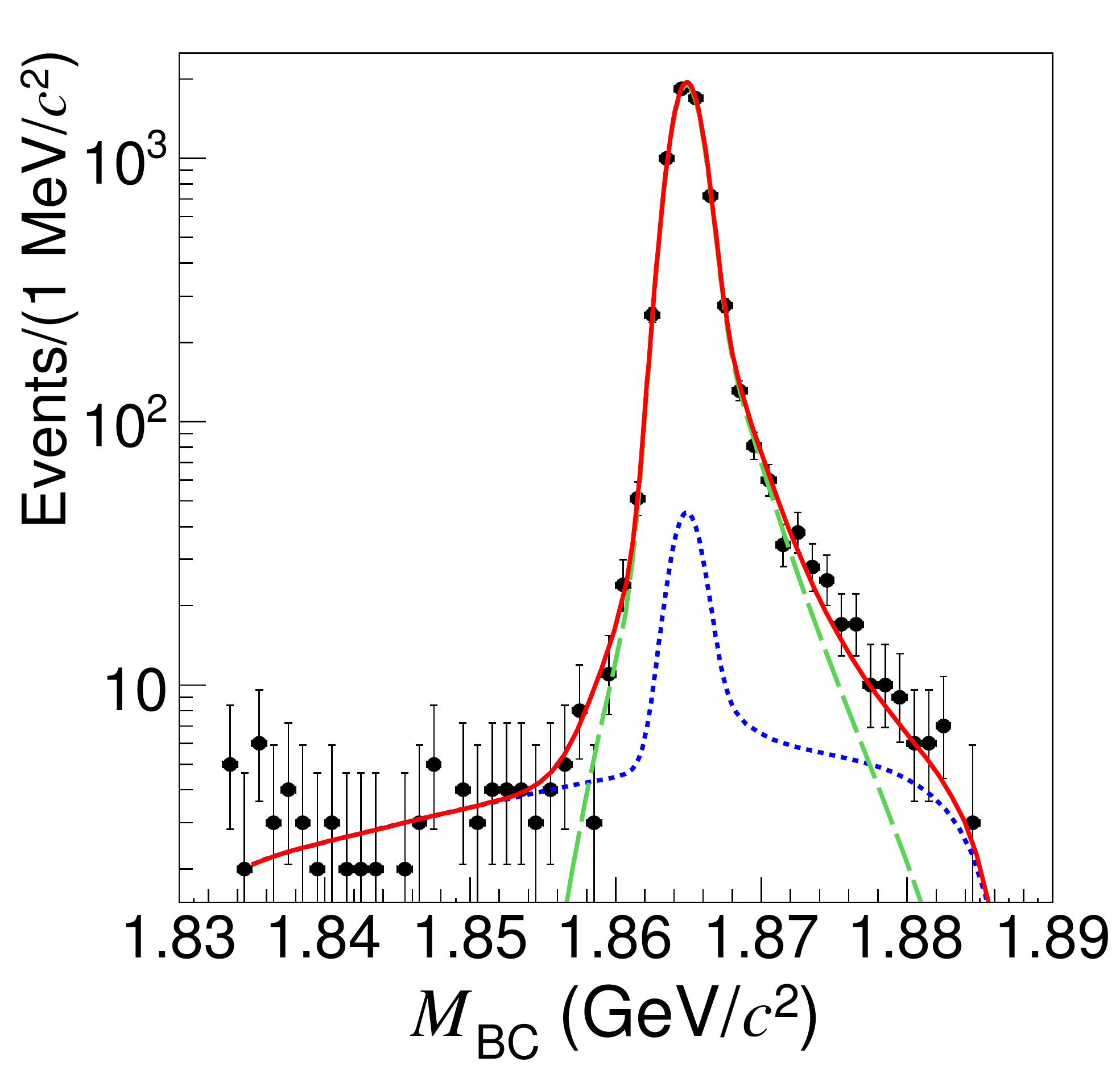}
\put(-25,120){(b)}
\end{minipage}
\begin{minipage}[b]{.62\columnwidth}
\epsfig{width=0.98\textwidth,file=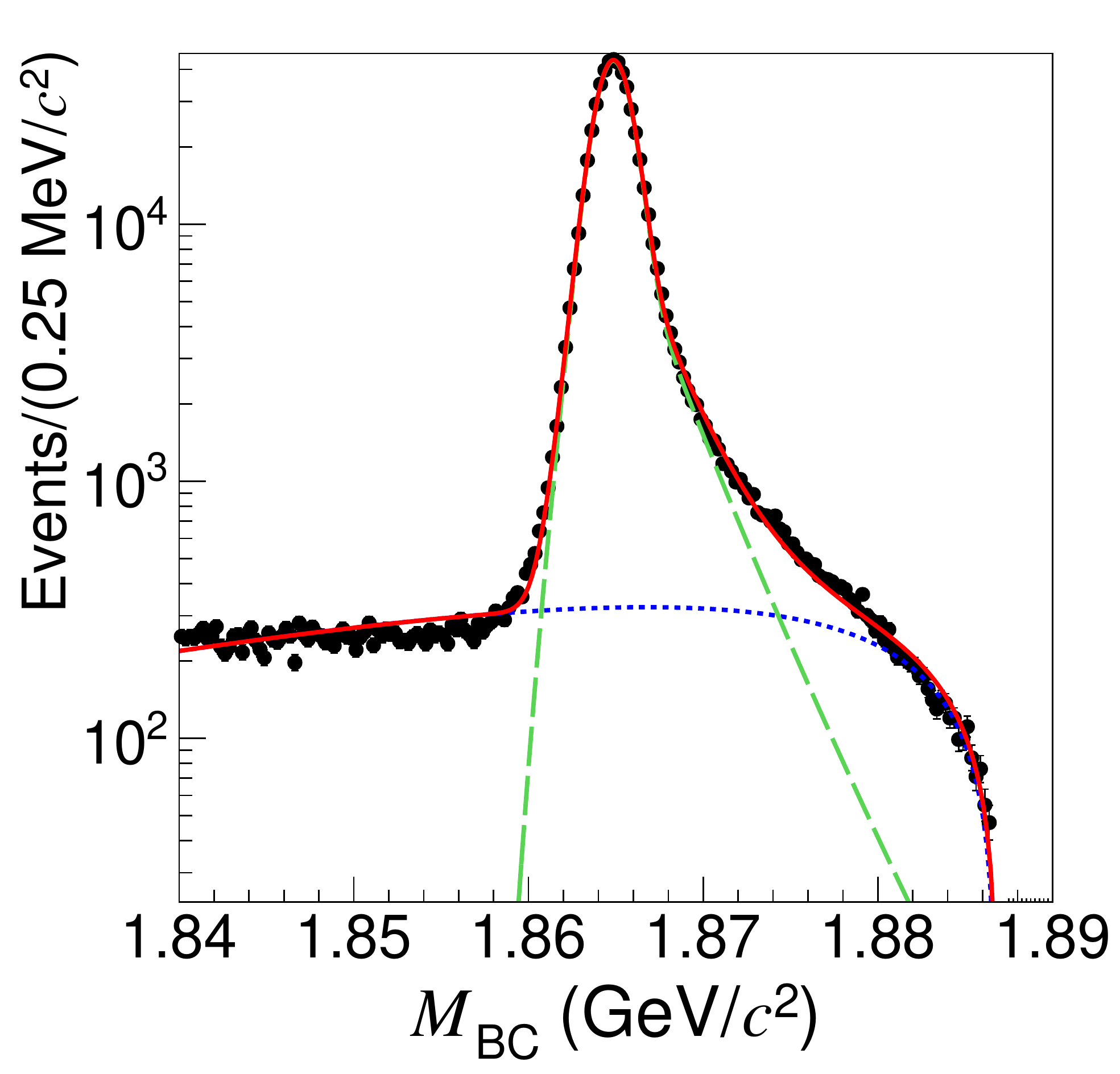}
\put(-25,120){(c)}
\end{minipage}
\caption{Fits to the $M_{\text{BC}}$ distributions of the DT of the data sample projected to the (a) signal side ($K^-\pi^+\pi^0\pi^0$) 
and the (b) tag side ($K^+\pi^-$) and fit to (c) the $M_{\text{BC}}$ distributions of the ST of the data sample, where the (black) dots with error bars are data, 
the (red) solid lines are the total fit, the (green) dashed lines are the signal, and (blue) dotted lines are the background.} \label{fig:data_yields_fit}
\end{center}
\end{figure*}

An updated MC sample based on 
our PWA results, called the PWA MC sample, is used to determine the efficiency. The PWA MC sample 
is the generic MC sample with the $K^-\pi^+\pi^0\pi^0$ versus $K^+\pi^-$ events replaced by the PWA signal
MC sample. 
All event selection criteria mentioned in Section~\ref{chap:event_selection} are 
applied except the $M_{\text{BC}}$ requirements. 
The projections to the signal and the tag side of the fit to the $M_{\text{BC}}$ distributions of the DT of data are shown in 
Figs.~\ref{fig:data_yields_fit}(a) and (b), respectively.
The background peak in the projection to the signal (tag)
  side axis is caused by events with a correct signal 
  (tag) and a fake tag (signal). 
The fit to the $M_{\text{BC}}$ distribution of the ST of data is shown in Fig.~\ref{fig:data_yields_fit}(c),
where both the mean values of the Gaussian function and the CB function agree well with our expectation for the $D^0$ mass.
The ST and DT data yields are determined to be $534,581\pm 769$ and $6,101\pm 83$, respectively.  
The ST and DT efficiencies based on the PWA MC sample are $(66.01\pm 0.03)\%$ and $(8.39\pm 0.04)\%$,
respectively. 

We further take the differences in efficiencies for $\pi^0$ reconstruction, tracking, and PID between the data and 
the PWA MC sample into account. For these differences, we obtain weighted-average efficiency differences 
$(\varepsilon_{\text{data}}/\varepsilon_{\text{MC}}-1)$ of $-0.69\%$, $1.83\%$, and $0.22\%$ for $\pi^0$ reconstruction, kaon tracking,
and pion tracking, respectively, while that for PID is negligible. More details are discussed in Section~\ref{sec:BF_systematics}. 
This correction is applied to obtain the corrected DT efficiency to be $(8.50\pm 0.04)\%$.

\subsection{Result of Branching Fraction}
Inserting the values of the DT and ST data yields, the ST efficiency, and the corrected DT efficiency 
 into Eq.~(\ref{eq:DTag_BF}),
we determine the BF of the $K^-\pi^+\pi^0\pi^0$ decay,
${\cal B}(D^0\rightarrow K^-\pi^+\pi^0\pi^0) \,=\,                                                                                                                                                                
(8.86 \pm 0.13(\text{stat}) \pm 0.19(\text{syst}))\%$. 
The systematic uncertainties are discussed in Section~\ref{sec:BF_systematics}.

\section{Systematic Uncertainties}\label{sec:uncertainties}
The systematic uncertainties of the PWA and BF measurement are 
discussed in Sections~\ref{sec:PWA_systematics} and~\ref{sec:BF_systematics}, respectively.
\subsection{Uncertainties for Amplitude Analysis}\label{sec:PWA_systematics}
The systematic uncertainties for our amplitude analysis are studied in four categories:
amplitude model, background, experimental effects, and fit bias.
The contributions from the different categories to the systematic uncertainties for the FFs and phases
are given in Tables~\ref{tab:ff_sys} and~\ref{tab:phase_sys}, respectively.
The uncertainties of these categories are added in quadrature to obtain the total 
systematic uncertainties.

The effects of the amplitude model arise from three possible sources: the $K\pi$ $S$-wave 
model, the effective barrier radii, and the masses and widths of intermediate particles.
To determine the systematic uncertainties due to the $K\pi$ $S$-wave model, the fixed 
parameters of the model are varied according to the  BABAR measurement uncertainties 
\cite{ASTON1988493,PhysRevD.78.034023}, listed in Table~\ref{tab:Kpi_swave_par}.
The effective barrier radius $R$ is varied from $1.5$ to $4.5$~GeV$^{-1}$ for 
intermediate resonances, and from $3.0$ to $7.0$~GeV$^{-1}$ for the $D^{0}$. The masses
and widths of intermediate particles are perturbed according to their published 
uncertainties in the PDG. The consequent changes of fitting results are considered
as the systematic uncertainties inherent in the amplitude model.

The effects of background estimation are separated into non-peaking background, 
and peaking background. 
The uncertainties associated with non-peaking background are studied by widening the 
$M_{\text{BC}}$ and $\Delta E$ windows on the signal side
to increase the non-peaking background. 
The peaking background can be mostly removed by the $K^0_S$ mass veto. 
However, this veto is also a source of uncertainties.
Its uncertainty is studied by widening this veto from the nominal 
$M_{\pi^0\pi^0}\notin (0.458, 0.520)$~GeV/$c^2$
to 
$M_{\pi^0\pi^0}\notin (0.418, 0.542)$~GeV/$c^2$.

The experimental effects are related to the acceptance difference between data and MC sample
caused by $\pi^{0}$ reconstruction, tracking, and PID efficiencies, which weight the 
normalization of the signal PDF, Eq.~(\ref{eq:gamma_epsilon}). To estimate the uncertainties associated with the 
experimental effects, the amplitude fit is performed varying $\pi^0$ reconstruction, 
tracking and PID efficiencies according to their uncertainties, and the changes of the nominal results are taken as the systematic uncertainties.

The fit bias is tested with 200 pseudoexperiment 
samples generated based on the PWA model.
The distribution of each FF or phase is fitted with a Gaussian function.
The difference of the mean and the nominal value is considered as the uncertainty 
associated with fit bias.

\begin{table*}[ht]
 \caption{FF systematic uncertainties (in units of statistical standard deviations) for:                                              
(I) the amplitude model, (II) background, (III) experimental effects, and (IV) fit bias.
The total uncertainty is obtained by 
adding all contributions in quadrature.}\label{tab:ff_sys}
 \begin{center}
 \centering
\begin{tabular}{lccccc}
 \hline
\hline
{\bf Amplitude mode}&{\bf I}&{\bf II}&{\bf III}&{\bf IV} &{\bf Total} \\\toprule
 \hline
$D\rightarrow SS$&&&&& \\ 
$D\rightarrow (K^-\pi^+)_S(\pi^0\pi^0)_S$ &1.518&1.258&0.072&0.235&1.987 \\ 
$D\rightarrow (K^-\pi^0)_S(\pi^+\pi^0)_S$ &1.524&0.835&0.078&0.004&1.740 \\ 
 \hline
$D\rightarrow AP, A\rightarrow VP$&&&&& \\ 
$D\rightarrow K^-a_1(1260)^+,                                                                                                                                                                                  
\rho^+\pi^0[S]$                             &1.293&0.436&0.030&0.363&1.412 \\ 
$D\rightarrow K^-a_1(1260)^+,                                                                                                                                                                                  
\rho^+\pi^0[D]$                             &0.938&0.368&0.024&0.284&1.046 \\ 
$D\rightarrow K_1(1270)^-\pi^+,                                                                                                                                                                                
K^{*-}\pi^0[S]$                             &1.643&1.175&0.160&0.182&2.035 \\ 
$D\rightarrow K_1(1270)^0\pi^0,                                                                                                                                                                               
K^{*0}\pi^0[S]$                             &1.562&0.567&0.034&0.036&1.662 \\ 
$D\rightarrow K_1(1270)^0\pi^0,                                                                                                                                                                               
K^{*0}\pi^0[D]$                             &0.989&0.541&0.035&0.068&1.201 \\ 
$D\rightarrow K_1(1270)^0\pi^0,                                                                                                                                                                               
K^-\rho^+[S]$                               &0.713&0.221&0.098&0.172&0.772 \\ 
$D\rightarrow (K^{*-}\pi^0)_A\pi^+,                                                                                                                                                                           
K^{*-}\pi^0[S]$                             &1.253&1.254&0.076&0.237&1.790 \\ 
$D\rightarrow (K^{*0}\pi^0)_A\pi^0,                                                                                                                                                                           
K^{*0}\pi^0[S]$                             &1.145&0.524&0.022&0.162&1.278 \\ 
$D\rightarrow (K^{*0}\pi^0)_A\pi^0,                                                                                                                                                                          
K^{*0}\pi^0[D]$                             &0.865&1.468&0.052&0.106&1.708 \\ 
$D\rightarrow (\rho^+ K^-)_A\pi^0,                                                                                                                                                                            
K^-\rho^+[D]$                               &1.249&0.812&0.084&0.186&1.504 \\ 
 \hline
$D\rightarrow AP, A\rightarrow SP$&&&&& \\ 
$D\rightarrow ((K^-\pi^+)_S\pi^0)_A\pi^0$&1.377&0.372&0.102&0.164&1.439 \\ 
 \hline
$D\rightarrow VS$&&&&& \\ 
$D\rightarrow (K^-\pi^0)_S\rho^+$        &1.308&0.252&0.070&0.476&1.416 \\ 
$D\rightarrow K^{*-}(\pi^+\pi^0)_S$      &0.381&0.549&0.023&0.166&0.689 \\ 
$D\rightarrow K^{*0}(\pi^0\pi^0)_S$      &0.880&0.417&0.078&0.232&1.005 \\ 
 \hline
$D\rightarrow VP, V\rightarrow VP$&&&&& \\ 
$D\rightarrow (K^{*-}\pi^+)_V\pi^0$      &0.688&0.752&0.033&0.273&1.056 \\ 
 \hline
$D\rightarrow VV$&&&&& \\ 
$D\rightarrow K^{*-}\rho^+[S]$          &0.980&1.354&0.059&0.371&1.713 \\ 
$D\rightarrow K^{*-}\rho^+[P]$          &0.425&0.506&0.031&0.348&0.747 \\ 
$D\rightarrow K^{*-}\rho^+[D]$          &1.365&0.598&0.049&0.398&1.543 \\ 
$D\rightarrow (K^-\pi^0)_V\rho^+[P]$   &0.695&1.223&0.027&0.140&1.414 \\ 
$D\rightarrow (K^-\pi^0)_V\rho^+[D]$   &1.335&0.848&0.237&0.401&1.649 \\ 
$D\rightarrow K^{*-}(\pi^+\pi^0)_V[D]$ &0.751&0.894&0.049&0.074&1.171 \\ 
$D\rightarrow                                                                                                                                                                                              
(K^-\pi^0)_V(\pi^+\pi^0)_V[S]$                 &0.818&0.443&0.046&0.211&0.955 \\ 
 \hline
$D\rightarrow TS$&&&&& \\ 
$D\rightarrow (K^-\pi^+)_S(\pi^0\pi^0)_T$&1.171&0.936&0.084&0.273&1.528 \\ 
$D\rightarrow (K^-\pi^0)_S(\pi^+\pi^0)_T$&0.803&0.188&0.068&0.018&0.828 \\ 
\hline
\hline 
\end{tabular}

 \end{center}
\end{table*}

\begin{table*}[ht]
 \caption{Phase, $\phi$, systematic uncertainties (in units of statistical standard deviations) for:    
(I) the amplitude model, (II) background, (III) experimental effects, and (IV) fit bias.
The total uncertainty is obtained by 
adding all contributions in quadrature.}\label{tab:phase_sys}
 \begin{center}
 \centering
\begin{tabular}{lccccc}
 \hline
{\bf Amplitude mode}&{\bf I}&{\bf II}&{\bf III} &{\bf IV}&{\bf Total} \\\toprule
 \hline
$D\rightarrow SS$&&&&& \\ 
$D\rightarrow (K^-\pi^+)_S(\pi^0\pi^0)_S$ &3.137&0.093&0.043&0.030&3.139 \\ 
$D\rightarrow (K^-\pi^0)_S(\pi^+\pi^0)_S$ &2.330&0.850&0.044&0.109&2.483 \\ 
 \hline
$D\rightarrow AP, A\rightarrow VP$&&&&& \\ 
$D\rightarrow K^-a_1(1260)^+,                                                                                                                                                                                  
\rho^+\pi^0[S]$                             &0.000&0.000&0.000&0.000&0.000 \\ 
$D\rightarrow K^-a_1(1260)^+,                                                                                                                                                                                  
\rho^+\pi^0[D]$                             &1.194&0.761&0.081&0.479&1.497 \\ 
$D\rightarrow K_1(1270)^-\pi^+,                                                                                                                                                                                
K^{*-}\pi^0[S]$                             &0.953&0.820&0.054&0.124&1.264 \\ 
$D\rightarrow K_1(1270)^0\pi^0,                                                                                                                                                                               
K^{*0}\pi^0[S]$                             &1.051&0.556&0.029&0.565&1.316 \\ 
$D\rightarrow K_1(1270)^0\pi^0,                                                                                                                                                                               
K^{*0}\pi^0[D]$                             &1.002&0.483&0.045&0.121&1.120 \\ 
$D\rightarrow K_1(1270)^0\pi^0,                                                                                                                                                                               
K^-\rho^+[S]$                               &2.007&0.188&0.079&0.847&2.188 \\ 
$D\rightarrow (K^{*-}\pi^0)_A\pi^+,                                                                                                                                                                           
K^{*-}\pi^0[S]$                             &1.208&0.706&0.048&0.455&1.472 \\ 
$D\rightarrow (K^{*0}\pi^0)_A\pi^0,                                                                                                                                                                           
K^{*0}\pi^0[S]$                             &1.711&0.365&0.053&0.214&1.750 \\ 
$D\rightarrow (K^{*0}\pi^0)_A\pi^0,                                                                                                                                                                           
K^{*0}\pi^0[D]$                             &1.501&0.605&0.051&0.187&1.630 \\ 
$D\rightarrow (\rho^+ K^-)_A\pi^0,                                                                                                                                                                            
K^-\rho^+[D]$                               &1.195&0.613&0.133&0.611&1.482 \\ 
 \hline
$D\rightarrow AP, A\rightarrow SP$&&&&& \\ 
$D\rightarrow ((K^-\pi^+)_S\pi^0)_A\pi^0$&2.039&0.410&0.045&0.446&2.127 \\ 
 \hline
$D\rightarrow VS$&&&&& \\ 
$D\rightarrow (K^-\pi^0)_S\rho^+$        &3.159&0.471&0.053&0.216&3.201 \\ 
$D\rightarrow K^{*-}(\pi^+\pi^0)_S$      &1.207&0.258&0.045&0.156&1.245 \\ 
$D\rightarrow K^{*0}(\pi^0\pi^0)_S$      &0.938&0.476&0.062&0.116&1.060 \\ 
 \hline
$D\rightarrow VP, V\rightarrow VP$&&&&& \\ 
$D\rightarrow (K^{*-}\pi^+)_V\pi^0$      &1.260&0.471&0.032&0.490&1.432 \\ 
 \hline
$D\rightarrow VV$&&&&& \\ 
$D\rightarrow K^{*-}\rho^+[S]$          &1.995&0.154&0.070&0.712&2.125 \\ 
$D\rightarrow K^{*-}\rho^+[P]$          &1.612&0.214&0.035&0.864&1.842 \\ 
$D\rightarrow K^{*-}\rho^+[D]$          &1.586&1.108&0.051&0.588&2.022 \\ 
$D\rightarrow (K^-\pi^0)_V\rho^+[P]$   &1.429&0.324&0.023&0.128&1.471 \\ 
$D\rightarrow (K^-\pi^0)_V\rho^+[D]$   &0.401&0.832&0.133&0.666&1.146 \\ 
$D\rightarrow K^{*-}(\pi^+\pi^0)_V[D]$ &1.445&1.313&0.040&0.190&1.962 \\ 
$D\rightarrow                                                                                                                                                                                              
(K^-\pi^0)_V(\pi^+\pi^0)_V[S]$                 &1.354&0.213&0.041&0.726&1.551 \\ 
 \hline
$D\rightarrow TS$&&&&& \\ 
$D\rightarrow (K^-\pi^+)_S(\pi^0\pi^0)_T$&2.544&0.724&0.057&0.189&2.653 \\ 
$D\rightarrow (K^-\pi^0)_S(\pi^+\pi^0)_T$&1.533&0.718&0.050&0.135&1.699 \\ 
\hline
\hline
 \end{tabular}
 \end{center}
\end{table*}

\subsection{Uncertainties for Branching Fraction}\label{sec:BF_systematics}
We examine the systematic uncertainties for the BF from the following 
sources: tag side efficiency, tracking, PID, and $\pi^0$ efficiencies for 
signal, $K^-\pi^+\pi^0\pi^0$ decay (PWA) model, yield fits, $K^0_S$ peaking background 
and the $K^0_S$ mass veto, and doubly Cabibbo-suppressed (DCS) decay.

The efficiency for reconstructing the tag side, $\bar{D}^0\rightarrow K^+\pi^-$, 
should almost cancel, and any residual effects caused by the tag side are expected 
to be negligible. 
Unlike the case of the tag side, the reconstruction efficiency of the signal side
does not cancel in the DT to ST ratio. This efficiency of the 
signal side is determined with the PWA signal MC sample. The mismatches of tracking, PID, 
and $\pi^0$ reconstruction between the data and MC samples, therefore, bring in 
systematic uncertainties.

One possible source of those uncertainties is that the momentum spectra simulated 
in the MC sample do not match those in data,
if there are any variations in efficiency versus momentum. This effect, however, is 
expected to be small due to the PWA MC sample's successful modeling of the momentum 
spectra in data, as shown in Fig.~\ref{fig:PWA_fitting_projection}.
The major possible source of the $\pi^0$ reconstruction, tracking, and PID systematic 
uncertainties is that, although the momentum spectra in the MC sample and data follow each 
other well, the efficiency of the MC sample disagrees with that of data as a function of 
momentum.  
This disagreement results in taking a correctly weighted average of 
incorrect efficiencies. We have performed an efficiency correction and choose 
$0.6\%$, $0.5\%$, $0.3\%$, and $0.2\%$ as the systematic uncertainties for the $\pi^0$ 
reconstruction, kaon tracking, pion tracking, and kaon/pion PID, respectively.
The uncertainty of the $\pi^0$ reconstruction efficiency is investigated with 
the control sample of $D^0\rightarrow K^-\pi^+\pi^0$ decays and the uncertainties for charged
tracks and PID are determined using the control sample of $D^+\rightarrow K^-\pi^+\pi^+$ decays,
$D^0\rightarrow K^-\pi^+$ decays, and $D^0\rightarrow K^-\pi^+\pi^+\pi^-$ decays.

To estimate the systematic uncertainty caused by the imperfections of the decay 
model, 
we compare our PWA model to another PWA model which only includes amplitudes 
with significance larger than $5 \sigma$. The relative shift on efficiency is less than $0.5\%$.
We therefore assign $0.5\%$ as the systematic for the effect of any remaining decay modeling 
imperfections on efficiency.

To get the uncertainty of the yield fit, we change the nominal $\Delta E$ window to 
a wider one, $-0.05<\Delta E<0.03$~GeV,
and the change of the BF is considered as the associated uncertainty.

The $K^0_S$ mass veto can veto most $K^0_S$ 
peaking background and reduce it to be only $0.07\%$ of the total events. However, the peaking background 
simulation is not perfect and the $K^0_S$ mass veto also removes about 13$\%$ of the signal events.
Thus, we estimate the uncertainty by narrowing the veto from 
$M_{\pi^0\pi^0}\notin (0.458, 0.520)$~GeV/$c^2$
to 
$M_{\pi^0\pi^0}\notin (0.470, 0.510)$~GeV/$c^2$,
while the $K^0_S$ peaking background increases from 
$0.07\%$ to $0.15\%$ and the BF change is 0.18$\%$ of itself. We take 
this full shift as the corresponding uncertainty. 

The smooth ARGUS background level is about $1.0\%$ in the signal region. In addition, 
the 2-D $M_{\text{BC}}$ ($K^-\pi^+\pi^0\pi^0$) versus $M_{\text{BC}}$ ($K^+\pi^-$) fit works well for the background determination.
Thus, we believe the uncertainty of the background with such small size will be very small and
neglect it. 

Our tag and signal sides are required to have opposite-sign kaons. This means that our 
DT decays are either both CF or both DCS. 
These contributions can interfere with each other, with amplitude ratios that are 
approximately known, but with a priori unknown phase. The fractional size of the 
interference term varies between approximately 
$\pm 2 |A_{\text{DCS}}/A_{\text{CF}}|^2 \simeq \pm 2 \tan^4 \theta_C$,
where $\theta_C$ is the Cabibbo angle
(the square in the first term arises as one power from each decay mode in the cross-term).
The two amplitude ratios are not exactly equal to $\tan^2\theta_C$, due to differing 
structure in the CF and DCS decay modes, but nonetheless we believe $2 \tan^4 \theta_C$ 
is a conservative uncertainty to set as an approximate ``$1 \sigma$'' scale to combine 
with other uncertainties.

Systematic uncertainties on the BF are summarized in Table~\ref{tab:BF_syst},
where the total uncertainty is calculated by a quadrature sum of individual contributions.


\begin{table}[tbh]
  \caption{$D^0\rightarrow K^-\pi^+\pi^0\pi^0$ BF systematic uncertainties. The total uncertainty is obtained by 
adding all contributions in quadrature.}
 \begin{center}
 \centering
 \begin{tabular}{lc}
\hline
\hline
Source & Systematic (\%)\\
\hline  
 Tracking efficiency  &    0.8 \\ 
 PID efficiency       &    0.4 \\ 
 $\pi^0$ efficiency   &    1.2  \\ 
\hline
 Decay model      &    0.5   \\  
\hline
 Yield fits (ST)  &    0.6   \\  
 Yield fits (DT)  &    1.2   \\  
\hline
 Peaking background &  0.2   \\  
\hline
 DCS decay correction  &    0.6   \\  
\hline  \hline
 Total            & 2.3 \\ 
\hline  \hline
\end{tabular}
 \end{center}
 \label{tab:BF_syst}
\end{table}

\section{Conclusion}
\label{CONLUSION}
%
Based on the $2.93$ fb$^{-1}$ sample of $e^+e^-$ annihilation data near $D\bar{D}$ 
threshold collected by the BESIII detectors, we report the first amplitude analysis of 
the $D^0\rightarrow K^-\pi^+\pi^0\pi^0$ decay and the first measurement of its decay branching 
fraction. We find that the $D^0\rightarrow K^-a_1(1260)^+$ decay is the dominant amplitude occupying $28\%$ of total FF (98.54\%) 
and other important amplitudes are $D\rightarrow K_1(1270)^-\pi^+$, $D\rightarrow (K^-\pi^0)_{S\text{-wave}}\rho^+$,
and $D\rightarrow K^{*-}\rho^+$, which is similar, in general, with the results of the BESIII $D^0\rightarrow K^-\pi^-\pi^+\pi^+$
amplitude analysis~\cite{PhysRevD.95.072010}.
Our PWA results are given in Table~\ref{tag:mode2_in}. 
With these results in hand, which provide access to an accurate 
efficiency for the $K^-\pi^+\pi^0\pi^0$ data sample,
we obtain ${\cal B}(D^0\rightarrow K^-\pi^+\pi^0\pi^0) \,=\, (8.86 \pm 0.13(\text{stat}) \pm 0.19(\text{syst}))\%$.


\begin{acknowledgements}
\label{sec:acknowledgement}
\vspace{-0.4cm}
The BESIII collaboration thanks the staff of BEPCII and the IHEP computing center for their strong support. This work is supported in part by National Key Basic Research Program of China under Contract No. 2015CB856700; National Natural Science Foundation of China (NSFC) under Contract No. 11835012; National Natural Science Foundation of China (NSFC) under Contracts Nos. 11475185, 11625523, 11635010, 11735014; the Chinese Academy of Sciences (CAS) Large-Scale Scientific Facility Program; Joint Large-Scale Scientific Facility Funds of the NSFC and CAS under Contracts Nos. U1532257, U1532258, U1732263, U1832207; CAS Key Research Program of Frontier Sciences under Contracts Nos. QYZDJ-SSW-SLH003, QYZDJ-SSW-SLH040; 100 Talents Program of CAS; INPAC and Shanghai Key Laboratory for Particle Physics and Cosmology; German Research Foundation DFG under Contract No. Collaborative Research Center CRC 1044; Istituto Nazionale di Fisica Nucleare, Italy; Koninklijke Nederlandse Akademie van Wetenschappen (KNAW) under Contract No. 530-4CDP03; Ministry of Development of Turkey under Contract No. DPT2006K-120470; National Science and Technology fund; The Knut and Alice Wallenberg Foundation (Sweden) under Contract No. 2016.0157; The Swedish Research Council; U. S. Department of Energy under Contracts Nos. DE-FG02-05ER41374, DE-SC-0010118, DE-SC-0012069; University of Groningen (RuG) and the Helmholtzzentrum fuer Schwerionenforschung GmbH (GSI), Darmstadt.
\end{acknowledgements}

\section{Appendix A: Amplitudes Tested}
\label{app:tested_modes}
The following is a list of all amplitude modes tested and found to have a significance smaller than $4\sigma$.  These are
not included in the final fit set.\\~\\
{\footnotesize $D\rightarrow PP, P\rightarrow VP$} \\ 
{\footnotesize
$D\rightarrow (K^{*-}\pi^0)_P\pi^+$\\ 
$D\rightarrow K^-(\rho^+\pi^0)_P$\\~\\
}
{\footnotesize \bf $D\rightarrow AP, A\rightarrow VP$} \\ 
{\footnotesize 
$D\rightarrow K_1(1270)^-\pi^+, K^{*-}\pi^0[D]$\\ 
$D\rightarrow K_1(1270)^0\pi^0, K^{*-}\pi^+[S]$\\ 
$D\rightarrow K_1(1270)^0\pi^0, K^{*-}\pi^+[D]$\\ 
$D\rightarrow K_1(1270)^0\pi^0, K^-\rho^+[D]$\\ 
$D\rightarrow K^-(\rho^+\pi^0)_A, \rho^+\pi^0[D]$\\
$D\rightarrow K^-(\rho^+\pi^0)_A[S]$\\ 
$D\rightarrow (K^{*-}\pi^+)_A\pi^0, K^{*-}\pi^+[S]$\\ 
$D\rightarrow (K^{*-}\pi^0)_A\pi^+, K^{*-}\pi^0[D]$\\ 
$D\rightarrow (K^{*-}\pi^+)_A\pi^0, K^{*-}\pi^+[D]$\\ 
$D\rightarrow (\rho^+ K^-)_A\pi^0,  K^-\rho^+[S]$\\~\\
}
{\footnotesize \bf $D\rightarrow AP, A\rightarrow SP$} \\
{\footnotesize
$D\rightarrow K^-((\pi^+\pi^0)_S\pi^0)_A$\\ 
$D\rightarrow K^-((\pi^0\pi^0)_S\pi^+)_A$\\ 
$D\rightarrow ((K^-\pi^0)_S\pi^+)_A\pi^0$\\ 
$D\rightarrow ((K^-\pi^0)_S\pi^0)_A\pi^+$\\ 
$D\rightarrow (K^-(\pi^+\pi^0)_S)_A\pi^0$\\ 
$D\rightarrow (K^-(\pi^0\pi^0)_S)_A\pi^+$\\~\\ 
}
{\footnotesize \bf $D\rightarrow VS$} \\
{\footnotesize 
$D\rightarrow (K^-\pi^0)_S(\pi^+\pi^0)_V$\\ 
$D\rightarrow (K^-\pi^+)_V(\pi^0\pi^0)_S$\\
$D\rightarrow (K^-\pi^0)_V(\pi^+\pi^0)_S$\\~\\
}
{\footnotesize \bf $D\rightarrow VP, V\rightarrow VP$} \\
{\footnotesize
$D\rightarrow (K^{*0}\pi^0)_V\pi^0$\\ 
$D\rightarrow (K^-\rho^+)_V\pi^0$\\~\\
}
{\footnotesize \bf $D\rightarrow VV$} \\ 
{\footnotesize 
$D[S]\rightarrow (K^-\pi^0)_V\rho^+ $\\
$D[S]\rightarrow K^{*-}(\pi^+\pi^0)_V $\\
$D[P]\rightarrow K^{*-}(\pi^+\pi^0)_V $\\
$D[P]\rightarrow (K^-\pi^0)_V(\pi^+\pi^0)_V$\\
$D[D]\rightarrow (K^-\pi^0)_V(\pi^+\pi^0)_V$\\~\\
}
{\footnotesize \bf $D\rightarrow TS$} \\
{\footnotesize
$D\rightarrow (K^-\pi^+)_T(\pi^0\pi^0)_S$\\ 
$D\rightarrow (K^-\pi^0)_T(\pi^+\pi^0)_S$\\~\\
}
{\footnotesize \bf Other}\\
{\footnotesize
$K^{*-}(1410)\pi^+$, $K^{*0}(1410)\pi^0$, $K^{*-}(1680)\pi^+$, $K^{*0}(1680)\pi^0$ \\ 
$K^{*-}_2(1430)\pi^+$, $K^{*0}_2(1430)\pi^+$, $K^{*-}_2(1770)\pi^+$, $K^{*0}_2(1770)\pi^+$ \\
$K^-a^+_2(1320)$ \\ 
$K^-\pi^+(1300)$ \\ 
$K^-\omega^+(1420)$ \\
$K^-a^+_1(1260)$ \\ 
$K^{*0}f_0(980)$\\ 
$K^{*0}_2(1430)(\pi^+\pi^-)_S$, $K^{*-}_2(1430)(\pi^+\pi^0)_S$\\ 
$K^{*-}_2(1430)\rho^+$\\ 
$K^{*0}_2(1430)f_2(1270)$\\ 
$(K^-\pi^+)_{S\text{-wave}}f_2(1270)$\\ 
$(K^{*-}\pi^+)_T\pi^0$, $(K^{*-}\pi^0)_T\pi^+$, $(K^{*0}\pi^0)_T\pi^0$\\ 
$(K^-\rho^+)_T\pi^0$\\ 
}


\begin{thebibliography}{99}
\bibitem{KLModes} Observation of a $K^0_Sn\pi$ mode alone suffices to infer 
the corresponding $\bar{K}^0n\pi$ final state, even when the partner 
$K^0_Ln\pi$ mode is not yet measured.
\end{thebibliography}
\end{document}